\documentclass[twocolumn,10pt]{article}
\usepackage[square]{natbib}
\usepackage[a4paper,margin=1in]{geometry} 
\usepackage{amsmath,amsfonts,amssymb}    
\usepackage{graphicx}                    
\usepackage{booktabs} 
\usepackage{chngcntr}
\usepackage{titlesec}                    
\usepackage{authblk} 
\usepackage{abstract}                    
\usepackage{cite}                        
\usepackage{float}                       
\usepackage{xcolor}                      

\newcommand{\exosims}[0]{{\tt EXoPLORE}}
\newcommand{\planeto}[0]{{HD\,189733\,b}}
\newcommand{\sysrem}[0]{{\tt SYSREM}}
\newcommand{\kms}[0]{km\,s$^{-1}$}
\titleformat{\section}{\bfseries\large}{\thesection.}{1em}{}
\titleformat{\subsection}{\bfseries}{\thesubsection.}{1em}{}

\usepackage{fancyhdr}
\title{\textbf{On the robustness of exoplanet atmospheric detections: insights from extensive simulations}}
\author[1]{A. S\'anchez-L\'opez}
\author[2]{Ana P. Mill\'an}
\affil[1]{Instituto de Astrof{\'i}sica de Andaluc{\'i}a (IAA-CSIC), Glorieta de la Astronom{\'i}a s/n, 18008 Granada, Spain}
\affil[2]{Institute ``Carlos I'' for Theoretical and Computational Physics, and Electromagnetism and Matter Physics Department, University of Granada, E-18071 Granada, Spain}
\date{\today} 

\begin{document}
\twocolumn[
\maketitle
\begin{onecolabstract}
\noindent The classical picture of our Solar System being the archetypal outcome of planet formation has been rendered obsolete by the astonishing diversity of extrasolar-system architectures.
From rare hot-Jupiters to abundant super-Earths and sub-Neptunes, most detected exoplanets have no analogs in our system, and their interior and atmospheric compositions remain largely unknown. Fortunately, new methodologies enable us to analyze exoplanet atmospheres, inferring their compositions, temperatures, dynamics, and even formation pathways. Specifically, ground-based high-resolution Doppler spectroscopy (HRDS) can disentangle spectral-line profiles of weak exo-atmospheric signals from the dominating features of Earth's atmosphere in the observed flux. 
For over a decade, HRDS has focused on hot Jupiters—close-orbiting gas giants—due to their high signal-to-noise ratio, which makes them ideal laboratories for advancing our knowledge. However, there have been concerns regarding potential biases in exo-atmospheric-detection methods, hindering comparative planetology.
Here we propose a modeling framework based on extensive simulations of HRDS exo-atmospheric observations to systematically explore \emph{in-silico} underlying biases in commonly-used pipelines, particularly under the presence of observational noise. Our findings show that exo-atmospheric detection-significances are highly contingent on details of the analysis-pipeline used, with different techniques responding differently to noise: A given technique may fail to recover a true-signal that is detected by another. 
Noise effects in the computed significances are non-trivial and pipeline-dependent. Statistical analyses provide a complementary tool to contextualize signal-significances, which will gain in relevance as we move towards studying the atmospheres of smaller, potentially habitable exoplanets with even weaker signals.
\end{onecolabstract}
\vspace{1em}
]

\section{Introduction}
Over the last two decades the field of exoplanet atmospheric research has grown exponentially, moving from original publications simply reporting exo-atmosphere detections, via the identification of atomic species \citep{charbonneau2002detection, vidal2003extended, redfield2008sodium, snellen2008ground}, to full retrievals of their physico-chemical and dynamical properties \citep{barstow2016consistent, madhusudhan2019exoplanetary, cont2022atmospheric, blain2023retrieval, holmberg2024possible, benneke2024jwst}. 
These studies are enabled by observing the primary transit or secondary eclipse of exoplanets. From an Earth's perspective, a primary transit occurs when the exoplanet passes in front of its host star, blocking a portion of the stellar disk. Further along the orbit, the exoplanet's illuminated side comes into view before and after being eclipsed by its stellar host. In both geometries, the exo-atmospheric spectral signatures (i.e. absorption, emission, or scattering) are imprinted on the stellar flux received. By comparing these observations with reference stellar data (i.e. out-of-transit or during the exoplanet's eclipse), where only the star's contribution is present, we reveal the exo-atmosphere signatures \citep{snellen2010orbital, brogi2012signature, birkby2013detection, nortmann2018ground, sanchez2019water}. 

Exo-atmospheric characterization can be performed both from space at low-to-mid resolution \citep{mccullough2014water, madhusudhan2014h2o, barstow2016consistent, benneke2024jwst} and from the ground at high resolution by applying the so-called high resolution Doppler spectroscopy (HRDS) technique \citep{snellen2010orbital, brogi2012signature, brogi2016rotation, brogi2018exoplanet, birkby2013detection, birkby2017discovery, nugroho2017high, nugroho2021first, nortmann2018ground, nortmann2024crires, sanchez2020discriminating, sanchez2022searching, stangret2020detection, kesseli2020search, kesseli2021confirmation, kesseli2022atomic, ehrenreich2020nightside, landman2021, blain2023retrieval, mansfield2024metallicity}. HRDS can resolve the relatively high and varying Doppler shift of the exo-atmospheric features, produced by the exoplanet's orbital motion, with respect to the nearly time-static spectral contributions of the Earth, of the exoplanet's host star, and of the instrumental effects.

HRDS involves different techniques for the removal of the Earth's and stellar atmospheric contributions (here referred to as ``preparation pipelines''), which are orders of magnitude larger than those of the exoplanet atmosphere. These techniques are typically based on applying principal component analysis (PCA) to subtract common modes from the data \citep[e.g.][]{birkby2013detection, birkby2017discovery, nugroho2017high, sanchez2019water}, or on the subtraction of polynomial fits to the time-evolution of the flux, hence tracing and removing the evolution of the Earth's transmittance, commonly referred to as ``telluric'' contribution \citep[e.g.][]{Brogi2019, blain2023retrieval}. 

For molecular compounds such as water vapor or methane, the buried exo-atmospheric signal can be further enhanced by performing a cross-correlation of the prepared data (i.e. telluric- and stellar-corrected) with a template of the expected exo-atmospheric absorption or emission \citep{snellen2010orbital, birkby2018exoplanet}. Thereby, the information of hundreds of ro-vibrational lines is merged in a cross-correlation function (CCF) peak that can indeed reveal these noise-level signals \citep{snellen2010orbital, brogi2012signature, brogi2016rotation, brogi2018exoplanet, birkby2013detection, birkby2017discovery, birkby2018exoplanet, nugroho2017high, nugroho2021first, hoeijmakers2018atomic, hoeijmakers2019spectral, hoeijmakers2020high, hawker2018evidence, alonso2019multiple, cabot2019robustness, sanchez2019water, sanchez2020discriminating, sanchez2022searching, ehrenreich2020nightside, merritt2020non, kesseli2020search, kesseli2021confirmation, kesseli2022atomic, cont2021detection, cont2022atmospheric, cont2022silicon, holmberg2022first, Cheverall2023}. 

In the case of a detection, the time-wise CCFs will show a trail of CCF peaks that closely follows the exoplanet velocities with respect to the Earth at each time frame, indicating the presence of the molecular compound in the exo-atmosphere. High-resolution spectroscopy also allows us to observe Doppler-shifts of the CCF trail with respect to the exoplanet's rest-frame velocities ($V_{rest}$), allowing us to study exo-atmospheric circulation \citep{snellen2010orbital, brogi2016rotation, sanchez2019water, sanchez2022searching, nortmann2024crires}.
Once the trace of a signal has been identified, there are a multitude of pathways to assess its significance, such as: (i) normalizing the CCFs by their standard deviation to compute the peak's signal-to-noise ratio \citep[S/N; see e.g.][]{alonso2019multiple, sanchez2019water}, (ii) performing a generalized Welch's t-test to compare two distributions of CCF values, one where we expect the exo-atmospheric signal (``in-trail'') and another ideally containing only noise \citep[``out-of-trail''; see e.g.][]{brogi2012signature, birkby2017discovery}, and (iii) the recently developed Bayesian retrievals for HRDS, coupled to a sampling algorithm of the multi-dimensional parameter-space, which allows us to perform a statistically-significant assessment of parameter values and uncertainties. Beyond, the retrieval's associated global evidences account for the goodness-of-fit and allow for model-comparison \citep[e.g.][]{Brogi2019, gibson2020detection, gibson2022relative, cont2022atmospheric, yan2023crires, lesjak2023retrieval, blain2023retrieval, maguire2024high}.

This diversity in data-preparation and signal-assessment pipelines originates from the wide variety of datasets explored. These include multiple instruments, different spectral regions (typically optical, near-infrared, or both), varying telluric conditions, and targeting exoplanet atmospheric signals that are not only orders of magnitude weaker than Earth's features but can also differ intrinsically among themselves. For instance, atomic spectral lines can be orders of magnitude stronger than their molecular counterparts. Interestingly, it has been shown that different analyses of the same dataset can draw very different pictures, depending on the strategies employed for data-preparation and signal-evaluation \citep{hawker2018evidence, cabot2019robustness, Cheverall2023}. 
In the best-case scenario, two different methodologies may yield similar detections, but further comparison of the inferred properties is hindered by the differing pipelines \citep{Cheverall2023}. 
In the worst, one avenue may find a robust detection (e.g. high Welch's $\sigma$-value) that is not confidently supported with another (e.g. a corresponding CCF peak with a low S/N) \citep{cabot2019robustness}. This issue is further complicated when considering the effect of employing different telluric and stellar correction strategies, since these involve non-linear operations whose interplay with the signal-detection and characterization pipelines is very hard or even impossible to assess \citep[see, e.g.][]{cabot2019robustness, Cheverall2023, blain2023retrieval}. 
The difference between reporting a weak hint of a signal or a confident detection may then originate in pipeline-dependent noise and residual structures that may align too well with what we expect to observe in the exoplanet's atmosphere. 
By the same token, optimization techniques for the aforementioned corrections may result in preferential selection of residuals that unwittingly inflate the detection's significance \citep{cabot2019robustness, Cheverall2023, blain2023retrieval}. This represents a significant issue hindering reproducibility and comparative planetology. 

Extensive simulations of exo-atmospheric observations can help us to systematically investigate the interplay between different sources of noise and pipeline details.
\emph{In-silico} experiments, in which the exo-atmosphere and noise are perfectly known, can become a fundamental tool to assess the robustness of potential exo-atmospheric signal-detections, while illustrating the core statistical phenomena underlying common techniques. 
This approach allows us to compare different data-preparation and signal-evaluation techniques, revealing their intrinsic biases.

In this paper we present {\tt EXoPLORE}, a computational framework to simulate high-resolution observations of exo-atmospheres (focused here on primary transits), prepare them for exo-atmospheric detections by removing the Earth's telluric contribution and the stellar features, and assess the significance of potential exo-atmospheric signals by means of different metrics. Through extensive simulations of HRDS observations based on a published detection of water vapor in the hot Jupiter HD\,189733\,b, we validate the framework by showing similarity between simulated and real results, and investigate differences arising from using either different correction strategies or significance-evaluation metrics. We find Bayesian retrievals being the most robust of common techniques against noise structures, being able to detect exo-atmospheric signals in datasets where other methods fail. This is at the expense of retrievals being also the most time and computationally demanding. Our study illustrates the building blocks of common techniques in the field, paving the way for future studies of more challenging, noisier signals from potentially habitable planets.

\section{Methods}
\label{sect:methods}
\subsection{High-resolution simulations}

From user-defined inputs describing the exoplanet, host star, and orbital properties \citep[see parameters used in Table\,1 of][]{blain2023retrieval}, a set of exposures is generated taking into account the specified exposure times, instrumental overheads, and the total observational time. Since our test case is a real dataset, we used the barycentric julian date and geometric airmass of HD\,189733's observations with CARMENES \citep{alonso2019multiple}, as provided in the data-headers by the standard pipeline {\tt caracal} v2.00 \citep{zechmeister2014flat, caballero2016carmenes}. This allowed us to ideally replicate the real scenario.

For each exposure, we obtain a model of the Earth's atmospheric transmittance from the ESO SKYCALC Tool \citep{noll2012atmospheric, jones2013advanced}, at a spectral resolving power of $10^6$, a reference geometric airmass $\mu$\,$=$\,$1.0$ and a precipitable water vapor content of $10$\,mm, which is close to the real value towards the host star at the Calar Alto Observatory during the observations. From this reference, $T_{{\oplus},\,0}$, the wavelength- and time-dependent telluric contribution $T_{{\oplus}}$ is computed as
\begin{equation}
\label{eq:tell_transm}
    T_{{\oplus}} \left( \lambda,\,t \right) = \exp \left( \mu(t)\, T_{{\oplus},\,0} \left( \lambda,\,t \right) \right).
\end{equation}

The model transmission spectrum of the exo-atmosphere $\mathcal{P}$ is computed using petitRADTRANS \citep{molliere2019petitradtrans}. We set a reference pressure P$_0$\,=\,10$^{3}$\,Pa for the annulus integration and the mass fractions $\text{X}_{\text{H}_2}$\,$=$\,0.75, $\text{X}_{\text{He}}$\,$\sim$\,$(12/37)\,\text{X}_{\text{H}_2}$\footnote{Following the He\,/\,H$_2$ ratio in Jupiter's atmosphere.}, and $\text{X}_{\text{H}_2\text{O}}$\,$=$\,3\,$\times10^{-4}$ \citep{boucher2021characterizing}. We note that there are significant differences between literature retrievals of the H$_2$O abundance, with some works finding higher values from high-resolution datasets, $\text{X}_{\text{H}_2\text{O}}$\,$\sim$\,$10^{-1.5}$ \citep{blain2023retrieval, klein2024atmospherix, finnerty2024atmospheric}. Here, we choose a conservative approach, using a commonly found value \citep[see e.g.][]{mccullough2014water, madhusudhan2014h2o, danielski20140, pinhas2019h2o, welbanks2019mass, boucher2021characterizing}, but  we note this is a subject of debate \citep{blain2023retrieval}. 
The chosen value yields high similarity between the results obtained for real and simulated data, whereas the higher abundance from \citet{blain2023retrieval, klein2024atmospherix, finnerty2024atmospheric} translates into an unrealistically strong detections. 
As for the thermal structure, it is set to a Guillot profile \citep{guillot2010radiative} with an atmospheric infrared opacity of $\kappa_{\text{IR}} =0.01$, an optical-to-infrared IR-opacity ratio $\gamma=0.4$, and a planet's equilibrium temperature of T$_{\text{equ}}$\,$=$\,$1170$\,K.  

The ingress and egress of the exo-atmospheric signal is modeled as a smooth transition of the exoplanet across the stellar disk. To do this, we used a model transit light curve computed with the {\tt BATMAN} package \citep{kreidberg2015batman2}, assuming a uniform limb-darkening model. 
Regarding the stellar spectrum $\mathcal{S}$, it is obtained from the PHOENIX database for an effective temperature of $5400$\,K and it is assumed to be constant for each exposure (i.e. no variability). 

Consecutively, the final exo-atmospheric and stellar models ($\mathcal{M}$\,$=$\,$\mathcal{P}$\,$\times$\,$\mathcal{S}$) are Doppler-shifted at each time-frame according to their velocity with respect to the Earth, following the equation
\begin{equation}
\label{eq:velocity_wrt_earth}
    v (t) = K\,\sin \left( 2 \pi \phi (t) \right) + v_{\text{sys}} - v_{\text{Bary}}(t),
\end{equation}
where $K$ is the projected reflex-velocity of the exoplanet and of the star towards the Earth, with values of $K_P$\,$=$\,$153$\,\kms\ and $K_s$\,$=$\,$0.21$\,\kms, respectively. Also in the equation, $\phi$ is the orbital phase of the planet (i.e. $0$ at mid-transit and 0.5 at mid-eclipse), $v_{\text{sys}}$\,=\,$-2.36$\,\kms\ is the systemic velocity, and $v_{\text{Bary}}$ is the velocity correction term accounting for the Earth's velocity with respect to the solar system's barycenter (provided by {\tt caracal}).

The final simulated dataset $F_{\text{sim}}$ (see also Fig.\,\ref{fig:S1}) is obtained by adding noise to each spectral point. Starting from the real data-quality per pixel over time, a matrix of noise $\epsilon \left( \lambda,\,t \right)$ is produced by randomly drawing values from a $\mathcal{N}\left(0,\,\sigma_{\text{noise}} \right)$ distribution, where the standard deviation is obtained as:
\begin{equation}
    \centering
    \sigma_{\text{noise}}\left( \lambda,\,t \right) = \frac{1}{S/N \left( \lambda,\,t \right)}. 
\end{equation}
Hence, the noisy observations can be calculated as follows:
\begin{equation}
    \centering
     F_{\text{sim}} \left( \lambda,\,t \right) = (T_{{\oplus}} \times \mathcal{M}) \left( \lambda,\,t \right)+ \epsilon \left( \lambda,\,t \right).
\end{equation}

\subsection{Preparing pipelines}
The preparing pipelines aim at cleaning the data from quasi-static Earth's, stellar, and instrumental spectral contributions so as to enable the detection of exo-atmospheric signals. In this paper we used two common approaches:
\begin{enumerate}
    \item The first approach is outlined in \citet{Brogi2019} and it is referred to here as ``BL19''. It consists in fitting each exposure with the mean observed spectrum using a second-order polynomial and dividing the data by the fit. Consecutively, the time-evolution of the residual flux in each pixel is modeled with a second-order polynomial, and the fit is divided from the data. This method effectively removes static spectral contributions, while preserving the exoplanet signal.
    \item The second avenue follows literature strategies \citep{birkby2017discovery, sanchez2019water, sanchez2020discriminating, sanchez2022searching, cont2021detection, cont2022atmospheric, nugroho2017high, nugroho2021first} and consists in a normalization step followed by the application of the {\tt SYSREM} algorithm \citep{tamuz2005correcting, mazeh2007transiting}, a PCA-based function that allows for proper weighting of each spectral pixel by its uncertainty \citep[see also full description of application in Appendix\,B of ][]{blain2023retrieval}. Here, we normalized the spectral matrices by fitting a quadratic polynomial to the pseudo-continuum of each spectrum in each spectral order. {\tt SYSREM} performs linear fits on the time axis of each pixel that closely approximate the almost wavelength-invariant evolution of the telluric and stellar contributions. However, the latter are not necessarily linear and several {\tt SYSREM} passes are usually required to successfully remove most of these features. In this work, we applied five {\tt SYSREM} passes to all spectral orders (except in Sect.\,\ref{sect:bias_opt} as it is described in the main text). In each pass, a model of the input spectral matrix is produced and then subtracted from the original input.
\end{enumerate}
    
For both pipelines, we place a telluric mask for pixels where the absorption drops below $20$\,\% of the flux continuum, and an additional mask for noisy columns that deviate over three times the standard deviation of the full spectral matrix.

\subsection{Recovering exo-atmospheric signals}

We implement the three most common strategies in the literature for assessing the significance of exo-atmospheric signals potentially present in the residual matrices. Namely, the signal-to-noise ratio evaluation, the performance of a Welch's t-test between distributions of cross-correlation values, and optimized samplings of the relevant-parameters space coupled to likelihood-based Bayesian estimators. Whereas the first two techniques require the computation of cross-correlations, Bayesian retrievals can equivalently show a CCF term or not in the $\ln(\mathcal{L})$ function. 
In all cases, the models of the exo-atmospheric absorption are computed using petitRADTRANS. For direct CCF, we set the mass fractions of hydrogen and helium to $\text{X}_{\text{H}_2}$\,$=$\,0.75 and $\text{X}_{\text{He}}$\,$\sim$\,$(12/37)\,\text{X}_{\text{H}_2}$, respectively. For H$_2$O and in order to reproduce the real analysis of the \planeto\ dataset, we use the nominal mass fraction of \citet{alonso2019multiple}, $\text{X}_{\text{H}_2\text{O}}$\,$\sim$\,7.7\,$\times10^{-4}$ (a volume mixing ratio of $10^{-4}$). Regarding the p-T structure, we use the nominal pressure--temperature profile from \citet{alonso2019multiple}. All models are convolved with a gaussian kernel so as to match the spectral resolving power of the instrument. The CCF is performed in a nominal range from $-250$ to $250$\,\kms using two velocity steps. The common $1.3$\,\kms\ is our nominal case and it is roughly the mean velocity step between pixels in the CARMENES NIR channel. In addition, we also test a step of $3.2$\,\kms, which is approximately the channel's mean resolution element. We calculate the weighted cross correlation $\operatorname{Corr}_w(X, Y; v)$ as follows:


\begin{equation}
\label{eq:ccf}
\begin{aligned}
\operatorname{Corr}_w(X, Y; v) = \frac{\operatorname{Cov}_w(X, Y_v)}{\sqrt{\operatorname{Cov}_w(X, X) \operatorname{Cov}_w(Y_v, Y_v)}},
\end{aligned}
\end{equation}

where $X$ is the prepared data, $Y_v$ is the template, $v$ is the velocity used to Doppler-shift the template at each time frame following Eq.\,\ref{eq:velocity_wrt_earth}, and $w = 1 /  \sigma^2$ are the weights (with $\sigma$ representing the data uncertainties). The covariance is computed as:

\begin{equation}
\label{eq:covariance}
\begin{aligned}
\operatorname{Cov}_w(X, Y) = \frac{\sum_{i} w_i (X_i - \bar{X}_w)(Y_i - \bar{Y}_w)}{\sum_i w_i},
\end{aligned}
\end{equation}

where i loops over all wavelength bins and the weighted means \( \bar{X}_w \) and \( \bar{Y}_w \) are given by:

\begin{equation}
\label{eq:weighted_mean}
\begin{aligned}
\bar{X}_w = \frac{\sum_{i} w_i X_i}{\sum_i w_i}, \quad \bar{Y}_w = \frac{\sum_{i} w_i Y_i}{\sum_i w_i}.
\end{aligned}
\end{equation}

With this definition, $\operatorname{Corr}_w(X, Y; v)$ is defined between $\pm 1$ and extreme values would mean perfect correlation ($+1$) or anticorrelation  ($-1$), while mean values peak at $0$. $\operatorname{Corr}_w(X, Y; v)$ is calculated for each spectral order of each exposure, hence obtaining a CCF matrix as a function of time and velocity-shift.\\

Next, these CCF matrices are summed for all spectral orders so as to merge all spectral information. In the resulting co-added matrix, we can evaluate the presence of signals as follows:

\begin{enumerate}
    \item Further enhancement of a potential exo-atmospheric signal is obtained by co-adding the order-summed CCF matrix over time in the exoplanet's rest-frame ($\text{V}_{\text{rest}}$). In this process, a range of $K_P$ from $-320$ to $320$\,\kms\ is explored for Eq.\,\ref{eq:velocity_wrt_earth} (i.e. $K_P$ can be measured). The result is a CCF map as a function of both $K_P$ and the rest-frame velocities (see examples in Fig.\,\ref{fig:Noise_CCF}). For each $K_P$, the S/N is calculated by dividing each CCF value by the CCF's standard deviation in a $\pm$\,$250$\,\kms\ interval, excluding $\pm$\,$25$\,\kms\ around each CCF value, which serves as a safety window. Ideally, the maximum-significance CCF-peak in the map should be located around the ground-truth $K_P$\,--\,$\text{V}_{\text{rest}}$ \citep[e.g.][]{brogi2018exoplanet, alonso2019multiple, sanchez2019water, sanchez2020discriminating, sanchez2022searching, landman2021, kesseli2022atomic}.
    \item We can divide the order-summed CCF matrix in two sets of CCF values: the ``in-trail'' distribution, containing a window CCF values across the expected exoplanet velocity trail (a nominal $5$-pixel-wide window); and the ``out-of-trail'' distribution, which includes the rest of the velocity bins. A safety window of $\pm$\,$25$\,\kms\ is also used around the in-trail window. The basis of this metric is assuming that the out-of-trail distribution contains uncorrelated CCF values that should be normally distributed with zero mean. On the contrary, the in-trail distribution, obtained around the correct $K_P$, should show a positive mean value. The assessment of similarity between both distributions is performed by means of a Welch’s t-test \citep{welch1947generalization}. The null hypothesis is that the in-trail and out-of-trail samples have the same mean and a drawn from the same parent distribution, which should be rejected with high t-values around the correct $K_P$. The associated p-values can then be converted into $\sigma$-values by calculating the quantile function (inverse of the cumulative distribution function) of the standard normal distribution for the given p-value \citep[e.g.][]{birkby2013detection, birkby2017discovery, nugroho2017high, brogi2018exoplanet, alonso2019multiple, sanchez2019water}. This approach provides a measure of how many standard deviations the observed t-statistic is from the mean under the null hypothesis. An example of a high-significance detection (significant deviation between distributions) is shown in Fig.\,\ref{fig:S8}.
    \item We performed a Bayesian retrieval on the residual matrices adopting the $\ln(\mathcal{L})$ formalism from \citet{blain2023retrieval}:
    \begin{equation}
        \ln(\mathcal{L}) = - \frac{1}{2} \sum \left( \frac{P_R (F) - P_R(M_{\theta})}{U_R} \right)^2, 
    \end{equation}
    where the function $P_R$ is the preparation pipeline applied to both the data $F$ and the model, $M_{\theta}$, where $\theta$ is the set of parameters used, $U_r$ are the propagated uncertainties of the data, and where we have dropped the wavelength and time dependencies for clarity. We note that the sum loops over these two variables.
    In Fig.\,\ref{fig:S9}, we show that this framework, in combination with the BL19 pipeline (used here), yields unbiased retrievals for noiseless data, thus showing its robustness. We set uniform priors for our parameters in the ranges of $-8$\,$<$\,$\log \left( \text{X}_{\text{H}_2\text{O}} \right)$\,$<$\,$0$, $90$\,$<$\,$K_P$\,$<$\,$250$\,\kms, $500$\,$<$\,T$_{eq}$\,$<$\,$2000$\,K, and $-25$\,$<$\,$\text{v}_{\text{rest}}$\,$<$\,$25$\,\kms, respectively. We sampled the parameter space using {\tt PyMultiNest}, a python-wrapper for the {\tt MultiNest} algorithm, using $100$ live points, the non-constant efficiency mode, an evidence tolerance of $0.5$, and a sampling efficiency of $0.8$ \citep{chubb2022exoplanet, blain2023retrieval}. {\tt PyMultiNest} returns the nested sampling global log-evidence, $\log (Z)$, which marks the logarithmic likelihood of the model given the data, integrating over all parameters. Hence, it provides a quantitative measure for model comparison.
\end{enumerate}

\subsubsection{Computational simulations of high-resolution observations}
\label{subsec:result_exoplore}

We have built a comprehensive, highly customizable simulator of high-resolution observations of exoplanet atmospheres: {\tt EXoPLORE} (EXOplanet Prediction Laboratory for Observations, Research, and Experiments).
{\tt EXoPLORE} comprises three modules: 
(i) data-simulation including all relevant spectral contributions (based on Blain et al.\,2024), namely the Earth's telluric transmittance, the stellar features, the exo-atmospheric absorption (or emission) and scattering, as well as the experimental noise (simulated to match the desired quality per pixel and exposure); 
(ii) data-preparation including normalization, and removal of the telluric and stellar components as described above; 
and (iii) data-analysis including the aforementioned techniques to evaluate the presence and significance of signals.

In this study we exploit \exosims\ to compare the performance of different data-preparing pipelines and assess the statistical significance of the results. We focus on the case of water vapor in the hot Jupiter \planeto, a well-known case study \citep{alonso2019multiple, sanchez2019water, sanchez2020discriminating, Cheverall2023, blain2023retrieval} that constitutes an ideal laboratory to test the performance of the tool. 

\section{Results and discussion}
\label{sect:stats_signal_detections}

\begin{figure*}[tbh!]
\centering
\includegraphics[width=120mm]{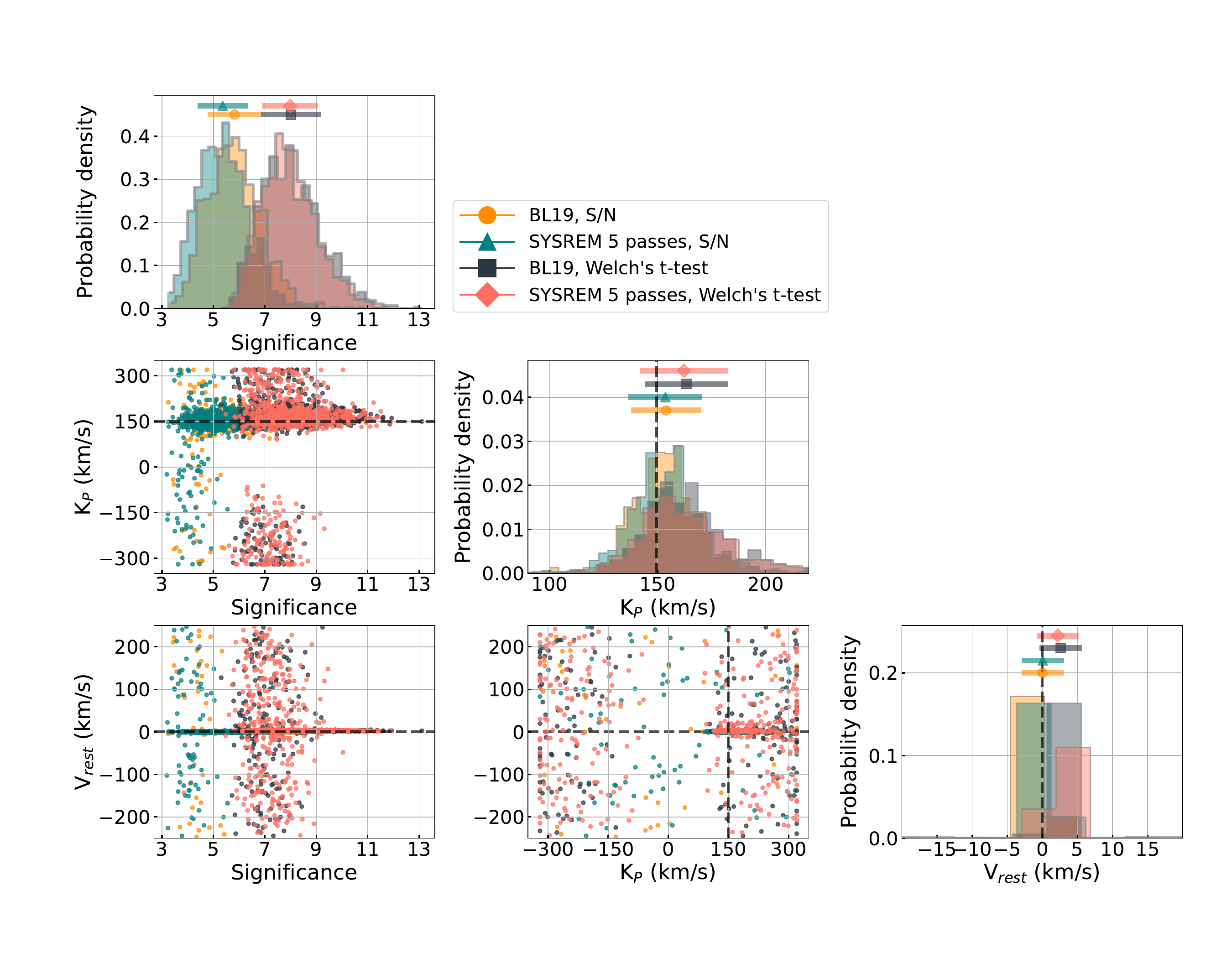}
\caption{Cross-correlation results of $1000$ \emph{in-silico} transit observations of \planeto, as they would appear with the CARMENES spectrograph. All datasets contain identical spectral contributions, with variations introduced only through additive noise matrices. Left to right, the diagonal plots display distributions of maximum-significance CCF signals, projected orbital velocities ($K_P$), and rest-frame velocities ($v_{rest}$), with horizontal error bars indicating the mean and 1-$\sigma$ uncertainties for each pipeline combination. $K_P$ and $v_{rest}$ distributions are shown in a zoomed-in range for clarity. The remaining panels show pair-wise parameter correlations, where each point represents a simulated observation. Grid steps are set at $1$\,\kms\ for $K_P$ and at the nominal CCF step of $1.3$\,\kms\ for $v_{rest}$ (the optimal CCF step of $3.2$\,\kms\ is shown in Fig.\,\ref{fig:S2}). Results are shown for the two preparation pipelines ({\tt SYSREM} and BL19) and two significance metrics (S/N and Welch’s $\sigma$-values). Dashed lines indicate the truth $K_P$\,--\,$V_{rest}$.
}
\label{fig:corner_plot}
\end{figure*}

\subsection{The case of water vapor in HD\,189733\,b}
\label{subsec:reproduction_H2O_HD189}
Water vapor is one of the most commonly found species in exoplanet atmospheres due to its dense forest of ro-vibrational absorption lines at NIR wavelengths. 
This property makes H$_2$O significantly easier to detect using CCF than other species. 
The H$_2$O absorption in the atmosphere of the archetypal hot Jupiter HD\,189733\,b has been investigated in multiple studies of both ground-based and space observations \citep{danielski20140, mccullough2014water, madhusudhan2014h2o, tsiaras2018population, brogi2018exoplanet, alonso2019multiple, sanchez2019water, sanchez2020discriminating, Cheverall2023, blain2023retrieval}. 
The Hubble Space telescope identified the absorption from individual H$_2$O bands \citep[e.g.][]{mccullough2014water, madhusudhan2014h2o, barstow2016consistent, tsiaras2018population}, which was then used to retrieve its abundance and  the exo-atmospheric temperature profile, whereas high-resolution studies reported evidence for supersonic high-altitude winds and even allowed for the inference of aerosol extinction \citet{birkby2013detection, brogi2018exoplanet, alonso2019multiple, sanchez2019water, sanchez2020discriminating, boucher2021characterizing, Cheverall2023, klein2024atmospherix, finnerty2024atmospheric}. 
Most recent HRDS studies of HD\,189733\,b have been performed using NIR transit observations obtained with the CARMENES spectrograph \citet{quirrenbach2016carmenes, quirrenbach2018carmenes}. 
CARMENES covers the NIR spectral region $0.96-1.71\,\mu$m over $28$ spectral orders, with a resolving power of $\mathcal{R}\sim82,400$. Specifically, we focus on the transit time-series observed on the night of $2017$ September $7$, consisting of $45$ exposures of $198$\,s. 

\subsubsection{Reproduction of the empirical H$_2$O detection}
We reproduced the water-vapor signal-detection in HD\,189733\,b's atmosphere using the CARMENES dataset, with the two data-preparation (\sysrem\ and BL19) and the two data-analysis (S/N and Welch's t-test) pipelines. 
As shown in Table\,\ref{Table:results_real_data}, we are able to reproduce the signal detections previously reported with the four pipelines. From these, the lowest significance is observed for the combination BL19\,--\,S/N, likely due the simpler nature of the BL19 pipeline, which is not able to fully remove the time-dependent contribution of telluric H$_2$O.
All pipelines find a CCF signal whose source changes its velocity with respect to thee Earth according to Eq.\,\ref{eq:velocity_wrt_earth}, with an exoplanet's projected orbital velocity ($K_P$) that is consistent with that of HD\,189733\,b.
Also, a significant blueshift of the H$_2$O signal is evidenced by an exoplanet rest-frame velocity (V$_{rest}$) of about $-$\,$5.2$\,\kms, which is well in agreement with previous reports of day-to-nightside wind in the upper atmosphere of this hot Jupiter. 

Our Welch's t-test studies of this real dataset find maximum signals away from the expected $K_P$\,--\,$\text{V}_{\text{rest}}$ and hence, we report the $\sigma$-values at the velocities of the maximum S/N signal in Table\,\ref{Table:results_real_data}. We find generally lower significances than the original study \citep{alonso2019multiple}, although they are still within $1$-$\sigma$. This is likely due to a combination of factors. For instance, the authors performed an in-depth analysis of CARMENES' line spread function (LSF) so as to accurately convolve their high-resolution models to match the wavelength-dependent resolving power of the instrument. This improves correlation with the true signal while reducing interference from tellurics. We did not perform such LSF characterization and thus our lower significances are expected.

\begin{table*}[t!]
\centering
\caption{Water vapor signal in \planeto's CARMENES dataset. Results obtained applying different preparation pipelines and significance-evaluation metrics.}
\begin{tabular}{lrrrrr}
Reference & Metric & Pipeline & Significance & $K_{\text{P}}$ & $\text{V}_{\text{rest}}$\\
\midrule
\citet{alonso2019multiple}  & S/N   & {\tt SYSREM} & $6.6$ & $160^{+45}_{-33}$ & $-3.9\pm1.3$\\
   & Welch's t-test$^\star$            & {\tt SYSREM} & $7.5$ & $160$ & $-3.9$ \\
\midrule
\midrule
This work & S/N & {\tt SYSREM} & $6.0$ & $140\pm40$ & $-5.2\pm2.6$ \\

 &    & BL19 & $4.6$ & $159^{+60}_{-44}$ & $-5.2\,\pm\,2.6$ \\
 & Welch's t-test$^\star$ & {\tt SYSREM} & $6.6$ & $-$ & $-$ \\
    
    &            & BL19 & $6.1$ & $-$ & $-$\\
\bottomrule
\end{tabular}\\
\footnotesize{$1$\,$\sigma$ error bars are indicated. A nominal CCF step of $1.3$\,\kms\ was used. $^\star$ Welch's significance quoted at the K$_{P}$\,--\,$\text{V}_{\text{rest}}$ of the maximum-S/N signal.}
\label{Table:results_real_data}

\end{table*}

\subsubsection{Statistical validation} We produced a sample of $1000$ synthetic observations of HD\,189733\,b's transit with the data-simulation module.
The time-dependent exoplanet absorption, Earth's telluric transmittance, and stellar spectral features were fixed for each observation, but random noise matrices $\epsilon_N \left( \lambda,t \right)$ were added for each realization. The noise matrices effectively degrade the data-quality until it matches the time- and wavelength-dependent signal-to-noise ratio of the real exposures ($\overline{S/N_{exp}}$). We illustrate this in Fig.\,\ref{fig:S1}, along with the effect of essential preparing steps in simulated datasets. The prepared datasets are consecutively analyzed with cross-correlation using the techniques discussed above.

In Fig.\,$\ref{fig:corner_plot}$ we show the results of this statistical analysis with a corner-plot of the significance (S/N and Welch's $\sigma$-value), $K_{\text{P}}$, and $\text{V}_{\text{rest}}$. 
The majority of simulated observations allow us to recover the maximum-significance CCF signal around the ground-truth $K_{\text{P}}$\,--\,$\text{V}_{\text{rest}}$. tuple for the four pipelines considered.
For S/N assessments, the expected significance ($<S/N>$) of the signal detection is $5.6 \pm 1.0$ for the BL19 pipeline and $5.4\pm 1.0$ for {\tt SYSREM}, in agreement within 1-$\sigma$ with the experimental measurements above. 
Similarly, the expected Welch's $\sigma$-value is $7.9\pm 1.1$ for BL19 and $8.0\pm 1.1$ for {\tt SYSREM}, also in line with earlier works \citep{alonso2019multiple}. Hence, our computational framework can reproduce within 1-$\sigma$ the significance of the published detection and, consequently, it can successfully predict the outcome of such real observations, although a fully dedicated analysis to this single dataset \citep[e.g.][]{alonso2019multiple, sanchez2019water} can reach higher significance detections.

Interestingly, we observe the Welch's t-test distributions show a significant shift towards higher significances, by a factor of $\sim$\,$1.5$, when compared to S/N results (see discussion in Sect.\,\ref{sect:reliability_metrics}). 
This has also been noted in empirical analyses in the past \citep{alonso2019multiple, cabot2019robustness, sanchez2019water}.
We also note that in the non-detection cases, the Welch's $\sigma$-values tend to cluster near the extreme $K_P$ values due to the telluric correction removing the non-Gaussian components of the data around $K_P$\,$=$\,$0$.
Thus, exploring wide intervals, also including the negative (unphysical) $K_P$-space, adds robustness to signal detection by helping to distinguish real signals from these cluster artifacts.
Regarding preparing pipelines, we do not observe significant differences on simulated datasets. Major differences in the results from Fig.\,\ref{fig:corner_plot} stem from different signal-evaluation metrics. This is most likely due to the absence of significant telluric variability or correlated-noise, which indeed play a key role in either hindering or inflating signals in real datasets.

\begin{figure*}[tbh!]
    \centering
    \includegraphics[width=1.5\columnwidth]{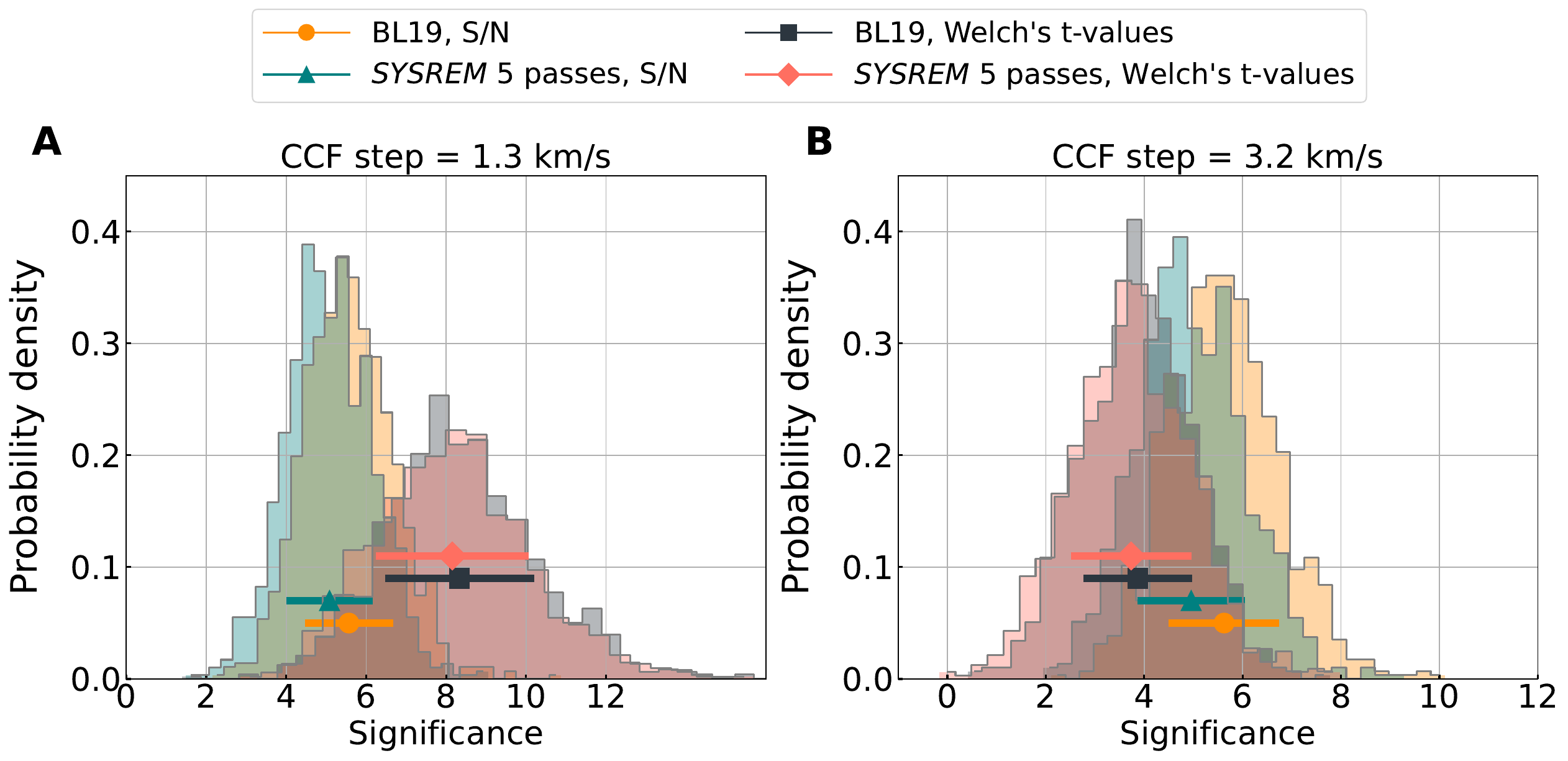}
    \caption{{Distributions of the CCF value measured exactly at the true $K_{\text{P}}$\,--\,$\text{V}_{\text{rest}}$ (i.e. the velocity-tuple where the exoplanet is simulated) in $1000$ \emph{in-silico} datasets for the two preparation pipelines ({\tt SYSREM} and BL19) and the two significance metrics (S/N and Welch's t-test). 
    Horizontal error bars indicate the mean and $1$-$\sigma$ uncertainties for each pipeline-metric combination. The cross-correlation analyses were computed using steps of $1.3$\,\kms\ (panel A) and $3.2$\,km/s (panel B).
    }
    }
    \label{fig:CCF_distributions}
\end{figure*}

Let us take a closer look at the implications of the spread observed in the significance distributions. The percentage of simulated observations that yield water vapor detections with significances within the aforementioned expected-value intervals is between $68\%$ and $70\%$ for all pipeline-combinations.
About $30\%$ of the \emph{in silico} observations will yield either a very high-significance detection (S/N\,$>$\,$6.6$) or, alternatively, weak signals (not even necessarily located at the ground-true $K_{\text{P}}$\,--\,$\text{V}_{\text{rest}}$) with which robust detections should not be claimed (S/N\,$<$\,$4.5$). In other words, nearly $15$\% of observations of an exo-atmosphere that presents H$_2$O absorption will yield non-detections or unrobust hints, whereas another $\sim$\,$15$\% of observations of the same atmosphere would show highly-significant detections, 
and this depends only on minute differences in their inherent observational-noise matrices $\epsilon_N \left( \lambda,t \right)$. 
We expand the investigation of this fundamental uncertainty in Sect.\,\ref{sect:source_of_var}.
 
We shift our focus now to the significance values at the ground-true $K_{\text{P}}$\,--\,$\text{V}_{\text{rest}}$ (instead of the maximum-significance signals studied before, wherever they may be located). That is, we study the significance exactly at the expected velocities of the simulated exoplanet with respect to the Earth (Fig.\,\ref{fig:CCF_distributions}).
In Fig.\,\ref{fig:CCF_distributions}A, we explore the results when using the nominal CCF step of $1.3$\,\kms\ \citep{alonso2019multiple, sanchez2019water, sanchez2020discriminating, sanchez2022searching, landman2021, rafi2024evidence}, which is the mean step-size between CARMENES' NIR pixels. This step-size defines the CCF-velocity resolution at which we evaluate the CCF peak’s Doppler shift. The distributions show similar averages, shapes, and shifts between S/N and Welch's t-test metrics than those observed in Fig.\,\ref{fig:corner_plot}. This is to be expected, since the maximum-significance signals were already predominantly found near the truth values. However, switching to a CCF step of $3.2$\,\kms\ (i.e. the mean resolution element of CARMENES' NIR channel; Fig.\,\ref{fig:CCF_distributions}B), we observe negligible changes for the S/N metric, but significant differences for Welch's t-test results, for which the previously-observed bias disappeared. As we will discuss in more detail in Sect.\,\ref{sect:reliability_metrics}, this is because the distribution of in-trail values (i.e. containing the exo-atmospheric signal) is oversampled for CCF steps lower than the instrumental resolution element.

\subsection{The reliability of signal evaluation metrics}
\label{sect:reliability_metrics}
As we showed in Sect.\,\ref{sect:stats_signal_detections}, the assessment of CCF-signal-significances and the assumed thresholds for detections can drastically change for different metrics. More importantly, the CCF velocity steps and ranges used for each metric can crucially impact the reported significance-numbers in the literature \citep[see also][]{cabot2019robustness, Cheverall2023}. In this section, we discuss this in detail for the same techniques employed above, which are among the most common in the field.

\subsection{The signal-to-noise ratio metric}

\begin{figure}[tbh!]
    \includegraphics[width=1.\columnwidth]{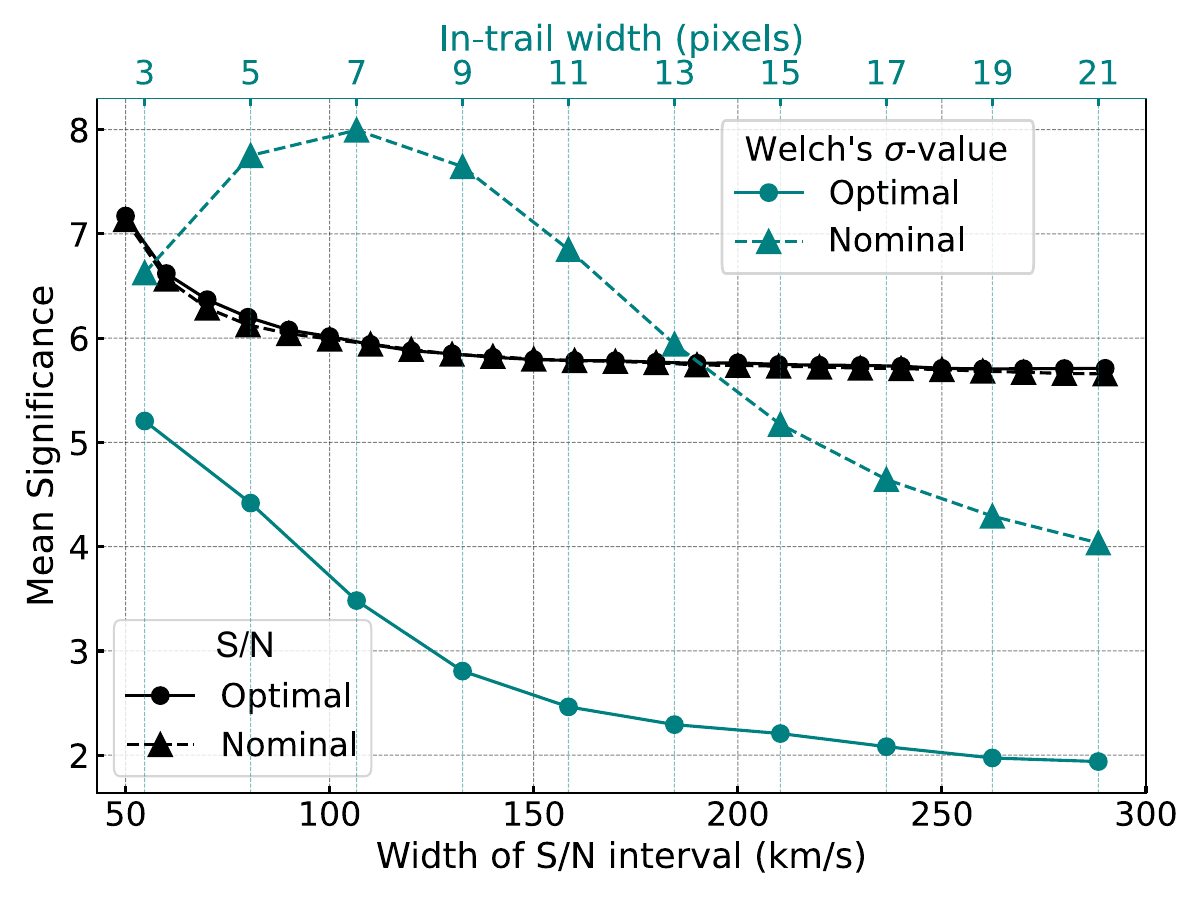}
    \caption{{Variability of significance-evaluation metrics with their respective calculation properties. The BL19 pipeline was used with the nominal CCF step ($1.3$\,\kms, dashed) and the optimal step ($3.2$\,\kms, solid), averaged over $300$ nights. 
    }
    }
    \label{fig:Sig_vs_Vinterval}
\end{figure}

The S/N evaluation is potentially the most straight forward assessment of a CCF signal's significance as it only implies dividing the CCF peak by the curve's standard deviation, but it also comes with a caveat. The resulting S/N can be heavily dependent on the velocity interval used to compute the standard deviation of the CCF noise, which is always a rather arbitrary choice \citep{Cheverall2023}. 
As reported in earlier studies of CARMENES' data for \planeto, the S/N appears to decrease for wider velocity intervals \citep{Cheverall2023}. Here, we computed the S/N in the nominal velocity range of $\pm$\,$250$\,\kms, which is a wide window in line with earlier works. In Fig.\,\ref{fig:Sig_vs_Vinterval} we confirm that the mean S/N value across realizations is indeed lower the wider the velocity interval used is. This decrease is the most pronounced from $50$ to $150$\,\kms, and wider velocity intervals will not significantly change the S/N computed.

\subsection{The Welch's t-test}

Several aspects stand out from the previous comparisons of the Welch's t-test and S/N metrics. For our nominal CCF step of $1.3$\,\kms\ the mean of the Welch's $\sigma$-values is significantly higher, about $8$ (Fig.\,\ref{fig:Sig_vs_Vinterval}), regardless of the preparing pipeline. In addition, the maximum Welch's $\sigma$-value is achieved for in-trail widths of $7$ pixels ($\pm$\,$3.9$\,\kms). However, this should not be interpreted as an intrinsic bias from Welch's t-test metric \citep[see also][]{cabot2019robustness, sanchez2019water}.
We find that this discrepancy is solely due to an oversampling of the in-trail distribution of CCF values when using a CCF sampling of $1.3$\,\kms. The mean resolution element of the CARMENES NIR channel, in velocity units, is approximately $3.2$\,\kms\ and hence, CCF calculations using a smaller velocity step \citep[see, e.g.,][]{nugroho2020searching, cont2021detection, cont2022atmospheric, cont2022silicon, dash2024constraints} incur in oversampling issues. As observed in Fig.\,\ref{fig:Sig_vs_Vinterval}, this is not relevant for the S/N metric, but the Welch's in-trail values are not independent from each other and hence, oversampling results in an overestimation of the points carrying signal. In fact, the bias-factor observed in Fig.\,\ref{fig:corner_plot} between Welch's t-test and S/N results is of $\sim$\,$1.5$, which is really close to the value of the oversampling correction factor $\sqrt{\text{Resolution}/\text{Sampling}}$ (i.e. effectively taking into account the number of independent CCF values). 

Regarding the dependency of the Welch's $\sigma$-values with the in-trail distribution's width (Fig.\,\ref{fig:Sig_vs_Vinterval}), we find the significances steadily decrease for wider in-trail windows. This is because too-wide in-trail widths dilute the significance by including noise (i.e. CCF points away from the exoplanet velocities with respect to the Earth), hence reducing the shift between the in-trail and out-of-trail distributions \citep[see also ][]{cabot2019robustness}. 
In Sect.\,\ref{subsec:reproduction_H2O_HD189} we used a nominal in-trail window of $5$ pixels to reproduce the methods of \citet{alonso2019multiple}, but we find our highest Welch's $\sigma$-values appear for the narrower in-trail width of $3$ pixels (i.e. centered at exactly the exoplanet velocities). 

Importantly, we find that the Welch's t-test is less reliable at finding exo-atmospheric signals when switching from the nominal (oversampled) to the optimal (adequately sampled) CCF step. That is, a larger spread in $K_P$ space and misplaced rest-frame velocities are observed for the latter by analyzing $1000$ \textit{in-silico} datasets with {\tt EXoPLORE} (Fig.\,\ref{fig:S2}). This suggests that, in some cases, oversampling the Welch's in-trail distribution of CCF values could be a reasonable approach to detect signals, considering the oversampling correction factor afterwards. 

It is thus crucial to properly contextualize the significance results quoted in publications, with detailed explanations of the methods employed and ranges considered, so that reproducibility and comparative-planetology studies can be carried out. Using a CCF step smaller than the instrumental resolution can inflate the reported significance of a CCF signal, with crucial implications for weak signals that are becoming our primary focus as we study smaller and colder exoplanets.
Therefore, the thresholds for atmospheric detections and what qualifies as high-significance are not necessarily consistent between the S/N and Welch's t-test metrics. Instead, they depend on arbitrary methodological choices, a fact that is insufficiently emphasized in the literature. 

\subsection{Bayesian retrievals}

In recent years, Bayesian retrievals have revolutionized HRDS, much as they previously did for space-based low-to-mid resolution data. These methods have opened new avenues for parameter estimation, as they enable the determination of temperature structures, atmospheric metallicity, elemental ratios, cloud pressure layers, among others, all with reliable uncertainty quantification \citep{Brogi2019, gibson2020detection, gibson2022relative, maguire2024high, yan2020lbt, cont2022atmospheric, blain2023retrieval, blain2024four}.
Here we explore whether Bayesian retrievals are also more resistant to  spectral noise. We used \emph{in-silico} datasets to retrieve the mass fraction of water vapor $\log \left( \text{X}_{\text{H}_2\text{O}} \right)$, the exoplanet's $K_{\text{P}}$, additional Doppler shifts observed in the exoplanet's rest-frame $\text{V}_{\text{rest}}$ (e.g. due to winds), and the equilibrium temperature (T$_{equ}$) of the exo-atmosphere, by using the log-likelihood function and retrieval scheme from \citet{blain2023retrieval} (see Sect.\,\ref{sect:methods}). Since performing retrievals in $1000$ datasets is not possible from a computational standpoint, we performed retrievals of a limited sample of $100$. 

\begin{figure}[tbh!]
    \centering
    \includegraphics[width=1.\columnwidth]{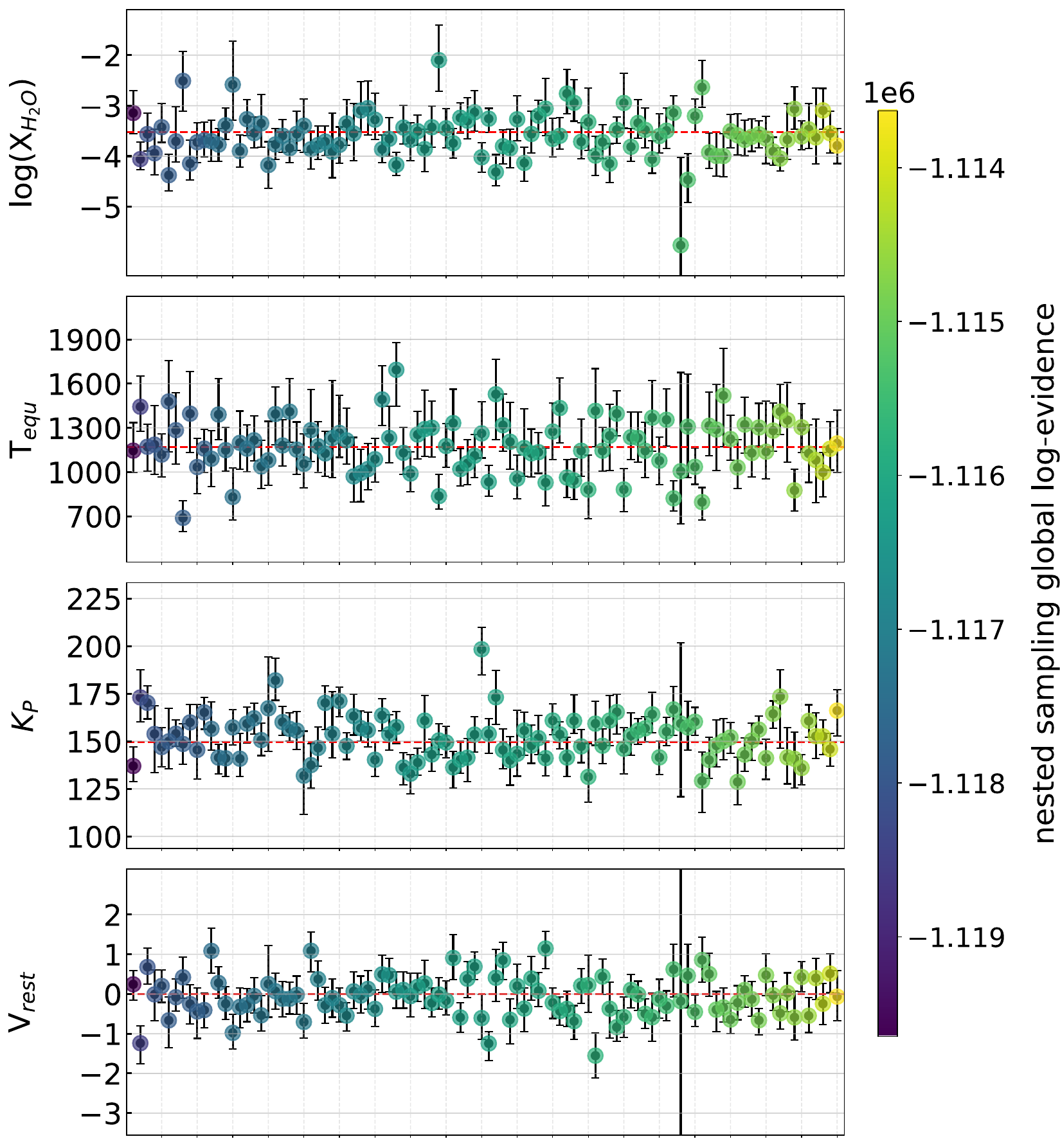}   
     \caption{{Results of Bayesian retrievals of $100$ in-silico observations of exo-atmospheres (nominal data quality), ordered by log-evidence (worst to best). From top to bottom, subplots show the retrieved parameters: H$_2$O mass fraction ($\log$$\left( \text{X}_{\text{H}_2\text{O}} \right)$), equilibrium temperature (T$_{eq}$), exoplanet’s $K_{\text{P}}$, and rest-frame velocity ($\text{V}_{\text{rest}}$). Each point represents the night's mean value with a $1$-$\sigma$ uncertainty. Red dashed lines denote the true values. 
     }
    }
    \label{fig:stats_retrievals}
\end{figure}

The retrieved parameters were found to cluster consistently near the truth values (Fig.\,\ref{fig:stats_retrievals}). Noise impacts retrievals by widening the uncertainty intervals of some parameters, with rare events of unconstrained values for which only upper limits are placed. However, the latter are frequently accompanied by other parameters still grouping their live points around the truth values (see also Fig.\,\ref{fig:S3}). Overall, retrievals seemingly suggest the presence of a real signal with a higher consistency than CCF-limited approaches. 
We note the effects of parameter degeneracy, which can be observed in the retrieval results as an anticorrelation between retrieved H$_2$O abundance and T$_{equ}$. Both parameters have a similar effect in the spectral lines, with deeper lines appearing for both high temperatures and high water abundances. Whereas the identification of the exo-atmospheric signal is still reliable, degenerate parameters cannot be accurately estimated.

\begin{figure}[tbh!]
    \centering
    \includegraphics[width=1\columnwidth]{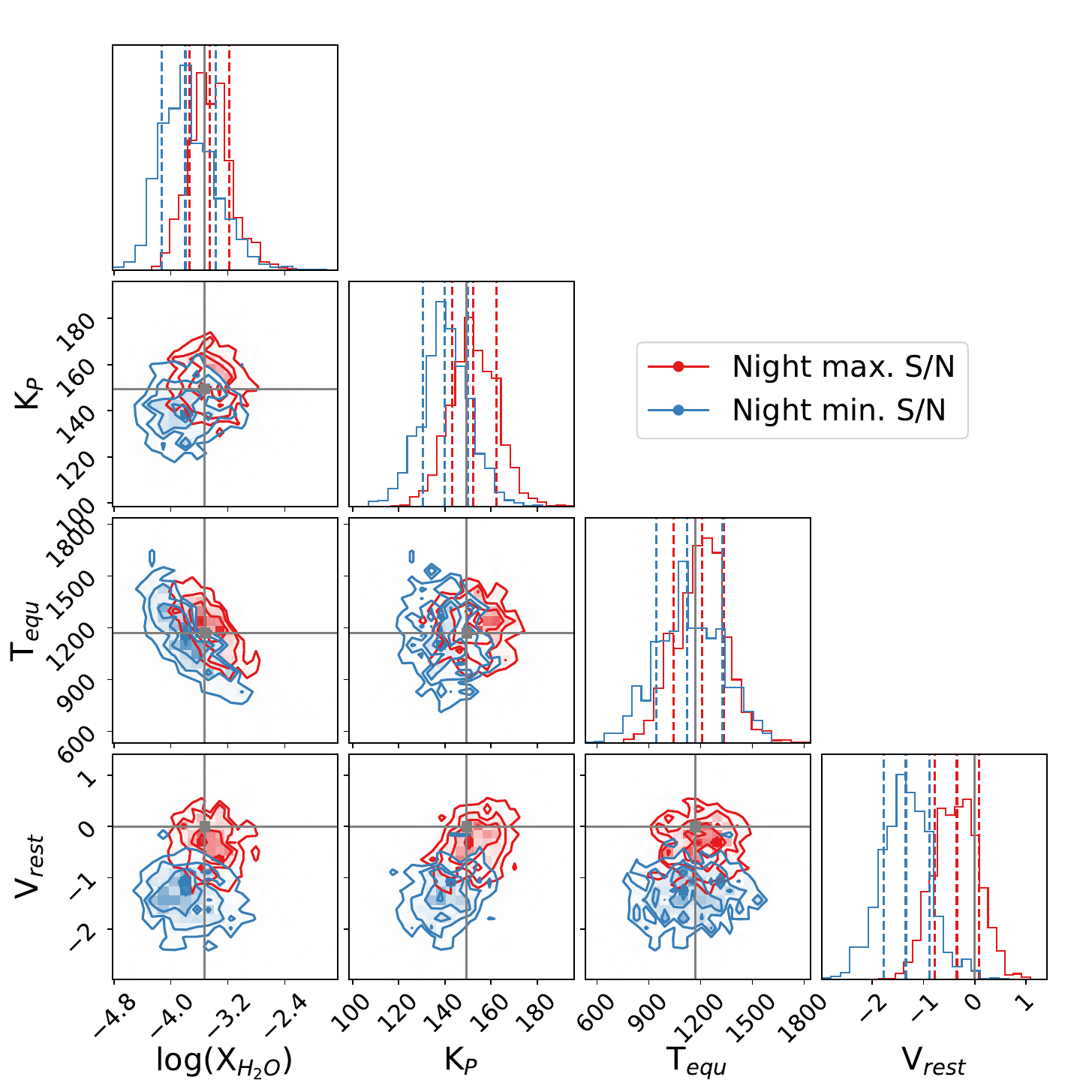}
     \caption{{Comparison of Bayesian retrievals of exo-atmospheric parameters in the nights with maximum (red) and minimum (blue) S/N signals (as per Fig.\,\ref{fig:corner_plot}). Dashed lines mark the $0.16$, $0.5$, and $0.84$ percentiles, while grey lines indicate truth values. 
     }
    }
    \label{fig:metric_comparisons}
\end{figure}

\subsection{Multi-metric analyses}
There is a direct, positive, and strong correlation between the computed significances with the S/N and Welch's t-test metrics (see Fig.\,\ref{fig:S4}). In other words, the \emph{in-silico} datasets showing the best and worst S/N are also most likely showing the highest and lowest Welch's $\sigma$-values, which is to be expected as they come from the same cross-correlation computation. Nevertheless, an exploration of both metrics is preferred to contextualize the significance-numbers reported, as discussed above.

By using the global evidence of the retrievals, we can also perform similar comparisons. We find that the highest global evidences are not found for the nights yielding the highest S/N and Welch's $\sigma$-values in standard CCF studies. We illustrate this lack of correlation in Fig.\,\ref{fig:metric_comparisons} (see Fig.\,\ref{fig:S4} for extended analyses), where we compare a retrieval of the best and worst simulated datasets as per their S/N significance (results from Fig.\,\ref{fig:corner_plot}). The Bayesian retrieval finds consistent water vapor signals in both extreme cases, with retrieved parameters in agreement with the truth values. The parameter estimation shows slightly larger uncertainties for the worst night, indicating the interference of H$_2$O-like noise, but the retrieval is indeed clearly able to recover signals in datasets for which common CCF\,--\,S/N strategies fail (blue retrieval). This poses strong implications, since previously discarded HRDS datasets could be revisited and potentially provide further insights on the observed exo-atmospheres. In any case, our results highlight the enhanced potential of retrieval analyses, in parallel with traditional CCFs studies, to obtain further and deeper insights into signal robustness and exo-atmospheric parameters.

\begin{figure*}[tbh!]
    \centering
    \includegraphics[width=1.5\columnwidth]{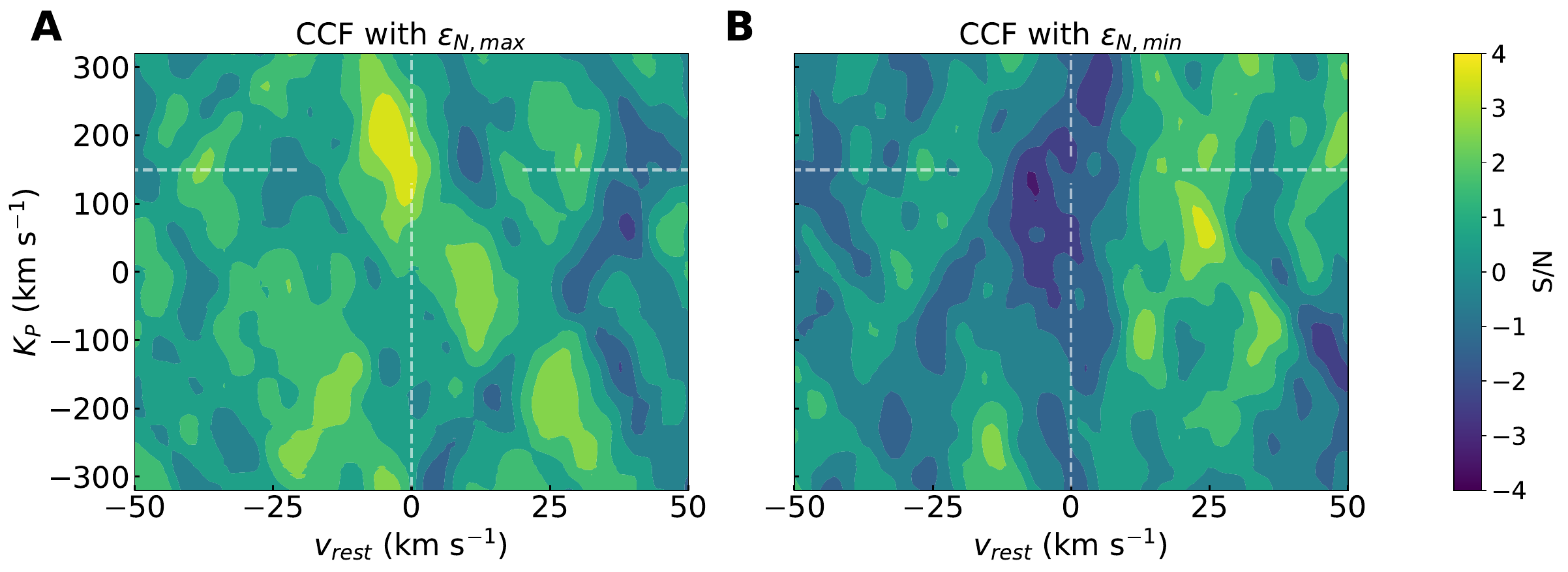}
    \caption{Cross-correlation maps of spurious H$_2$O-like signals shown in S/N units, plotted against exoplanet rest-frame velocity ($\text{V}_{\text{rest}}$, horizontal axis) and projected orbital velocity ($K_P$, vertical axis). These maps are derived from cross-correlation function (CCF) analyses between BL19-prepared noise matrices and the nominal H$_2$O template. The white lines mark the expected $K_P$ and $\text{V}_{\text{rest}}$ values of \planeto. Panels A and B display the results for $\epsilon_{N,, max}$ and $\epsilon_{N,, min}$, representing the spectral noise matrices of the nights yielding maximum and minimum S/N in Fig.,\ref{fig:corner_plot}, respectively. Lighter (yellow) regions denote strong correlations, while darker (dark blue) areas indicate anticorrelations, as shown by the colorbar. Minute differences in the noise matrices of two nearly identical nights can lead to either strong correlations (panel A) or anticorrelations (panel B) around the ground-truth $K_P$\,--\-,$v_{rest}$. Similar trends are observed in Welch's t-test analyses (not shown).
    }
    \label{fig:Noise_CCF}
\end{figure*}

\begin{figure*}[tbh!]
    \centering
     \includegraphics[width=1.\columnwidth]{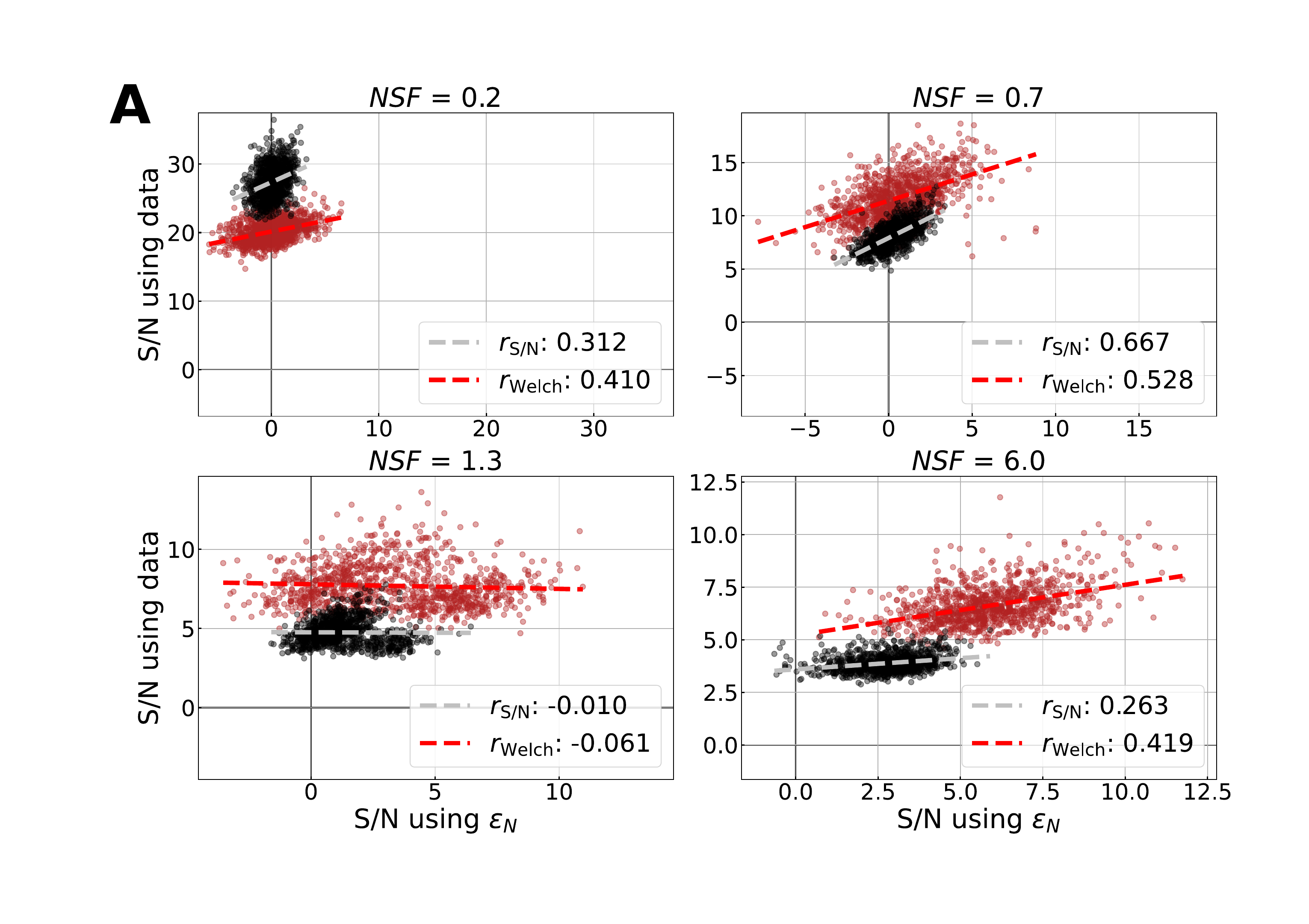}
     \includegraphics[width=1.\columnwidth]{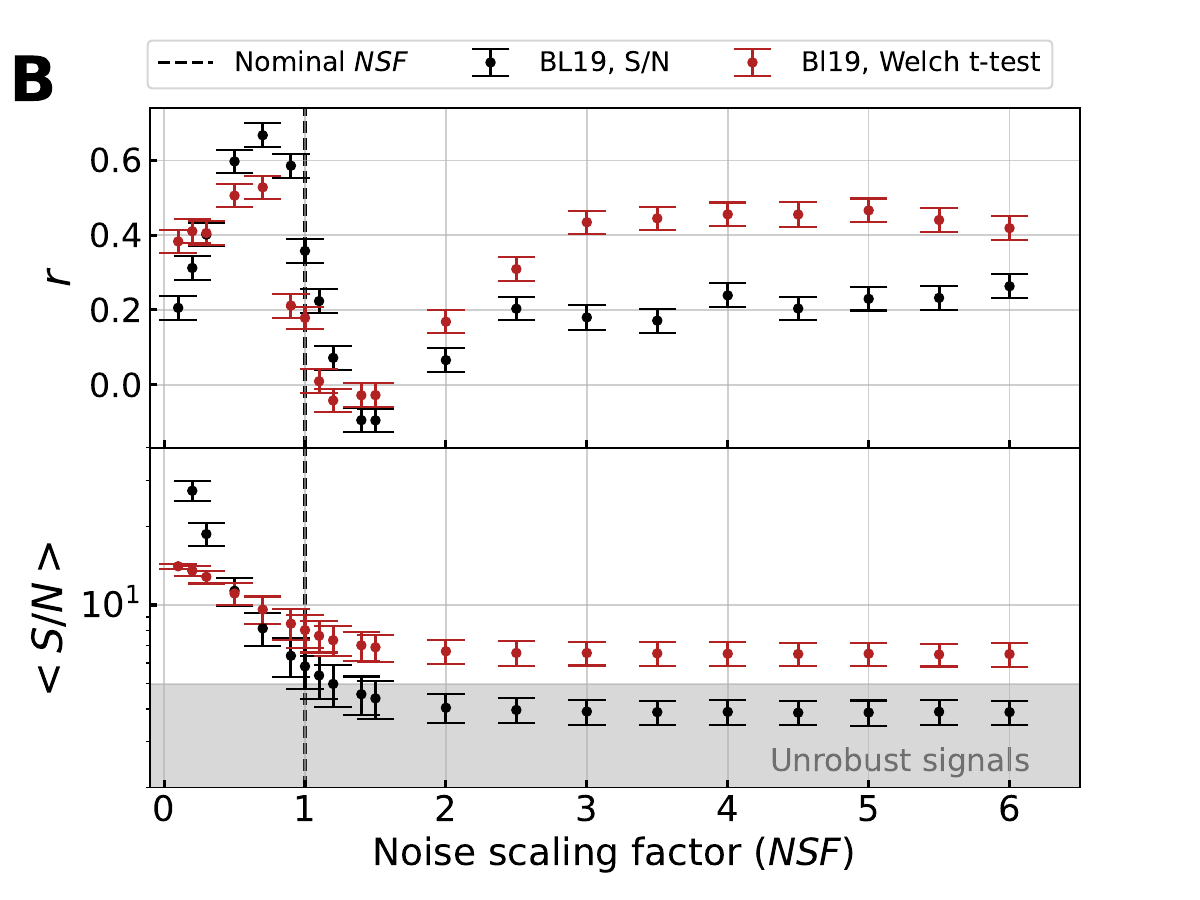}
    \hspace{0.01\textwidth} 
    \caption{Effect of noise on maximum-significance CCF signals from $1000$ simulated nights, in units of S/N (black) and Welch's t-values (dark red), for the BL19 preparation pipeline ({\tt SYSREM} results shown in Fig.\,\ref{fig:S7}). A CCF step of $1.3$\,\kms\ was used (similar results are observed for a $3.2$\,\kms\ step, Fig.\,\ref{fig:S6}). Panel A: Pearson correlation ($r$) between maximum-significance CCF peaks from in-silico data (vertical axis) and their noise matrices (horizontal axis) across varying \textrm{NSF} values. S/N and Welch's t-values are represented in black and dark red, respectively. Dashed lines show linear fits for S/N (grey) and Welch's metrics (red), with the associated $r$-correlations shown in the legends. Panel B: Pearson correlation coefficient as a function of the noise scaling factor (\textrm{NSF}) (top), and the mean maximum-significance values for each \textrm{NSF} (bottom). The vertical line marks the nominal uncertainties for real data (i.e., $\text{NSF}=1.0$). The shaded region in the lower plot marks unreliable signals below a threshold of $5$.
    }
    \label{fig:Stat_scale_factors}
\end{figure*}

\subsection{The source of intrinsic variability}
\label{sect:source_of_var}

To investigate the source of variability for each metric, we analyzed the noise matrices of the nights yielding the highest ($\epsilon_{N,\, max}$) and lowest ($\epsilon_{N,\, min}$) detection significances for each pipeline.
We treated these noise matrices as if they were actual observations and computed their cross-correlation with the H$_2$O template. The resulting CCF $K_P$\,--\,$\text{v}_{\text{rest}}$ maps are shown in Fig. $\ref{fig:Noise_CCF}$ for the extreme S/N cases (S/N$_{\text{min}}$\,$=$\,$3.2$ and S/N$_{\text{max}}$\,$=$\,$9.6$) observed in Fig.\,\ref{fig:corner_plot} when using the BL19 preparing pipeline. We found that the night $\epsilon_{N,\, max}$ yields a high-correlation region around the expected exoplanet's location, whereas a significant anticorrelation is observed in this region for $\epsilon_{N,\, min}$. That is, the random noise fluctuations may correlate or anticorrelate with the H$_2$O template, interfering with the inferred signal in a significant manner. This result may have relevant implications when interpreting weak signals (``hints'' at S/N\,$\sim$\,$4$), since it is impossible to know for certain how much the noise is acting to enhance or lower the detection significance when using a single observation. 

We can provide a more complete perspective of this issue by searching for H$_2$O in $1000$ \emph{in-silico} observations of a dry exo-atmosphere (no H$_2$O; Fig.\,\ref{fig:S5}). Random signals are spread with $<S/N>$\,$=$\,$3.9\pm 0.4$. High S/N values, above S/N of $4$ are quite likely ($33.4$\,\% of realizations; $8.1$\,\% for S/N\,$>$\,$4.5$), but the occurrence of noise-driven signals in a region around the true $K_P$\,--\,$\text{V}_{\text{rest}}$ and in the range of $3.5<$\,S/N\,$<5.5$ is extremely low ($8$ out of $1000$ events or $0.8\%$ of realizations). Therefore, it is the fact that the maximum-significance signal is detected around the expected $K_P$\,--\,$\text{V}_{\text{rest}}$ what adds robustness to the results, regardless of thresholds and CCF-significance values obtained by any method. This is particularly relevant because noise alone can produce CCF peaks with S/N\,$\sim$\,$5$ and Welch's $\sigma$ values of $9$ (for a CCF step of $1.3$\,\kms) or $5$ (for a CCF step of $3.2$\,\kms). This values are often interpreted as high-significance signals at any $K_P$\,--\,$\text{V}_{\text{rest}}$.


Noise interference in significance maps depends on data quality and, specifically, on the signal-to-noise ratio of the exposures ($\overline{S/N_{exp}}$). 
To quantify this effect, we measured the Pearson's correlation coefficient $r$ between the significance of the H$_2$O signal in $1000$ \textit{in-silico} nights and that of their corresponding noise matrices $\epsilon_N \left( \lambda,t \right)$. This entails computing the $K_P$\,--\,$\text{v}_{\text{rest}}$ significance maps, for both the full \emph{in-silico} datasets (including all spectral contributions and noise) and then for each dataset's noise matrix $\epsilon_N \left( \lambda,t \right)$ only. This results in $1000$ sets of paired observation--noise $K_P$\,--\,$\text{v}_{\text{rest}}$ maps. In Fig.\,\ref{fig:Stat_scale_factors}A we show the $r$-correlation between the maximum-significance signals in each paired map, which informs us of the importance of observational noise in the significance of potential detections. A more complete analysis is achieved by exploring a wide range of data qualities, simulated by modifying $\overline{S/N_{exp}}$. For simplicity, we quantify this by a noise scaling factor \textrm{NSF} that modifies $\overline{S/N_{exp}}$ by the nominal value of $150$, corresponding to the mean $\overline{S/N_{exp}}$ in the continuum of the real CARMENES' observations of \planeto\ \citep{alonso2019multiple}. 

We fully explore the $r$ dependence on \textrm{NSF} in Fig.\,\ref{fig:Stat_scale_factors}B (top panel). For both significance metrics, the $r$-correlation follows a non-monotonic trend, with a local maximum at low \textrm{NSF} ($0.7$ and $0.5$,
respectively for the S/N and Welch's t-test), followed by a common local minimum (\textrm{NSF}\,$=$\,$1.4$), and a final increase until a plateau is reached at \textrm{NSF}\,$\gtrsim$\,$3.5$.
Regarding the S/N of the detection (Fig.\,\ref{fig:Stat_scale_factors}B, bottom), it decays exponentially with the \textrm{NSF} (i.e., noisier datasets) and, from \textrm{NSF}\,$\geq$\,$1.2$, it drops below the detection threshold of $S/N$\,$=$\,$5$, as the \emph{in-silico} data become noise-dominated.
As hinted above, the higher basal significances found in the Welch's t-test show noise-dominated data presenting maximum significance signals that are still $\gtrsim$\,$6$. This is, however, solved for a CCF step that matches the resolution element of the instrument, bringing the mean Welch's t-test significances closer to the S/N ones (see Fig.\,\ref{fig:S6}).

Overall, we distinguish the following regimes in Fig.\,\ref{fig:Stat_scale_factors}B:
\begin{enumerate}
    \item For extremely high observation-quality (\textrm{NSF}\,$<$\,$0.5$, $\overline{S/N_{exp}}$\,$\gtrsim$\,$300$, top-left panel in Fig.\,\ref{fig:Stat_scale_factors}A), the exo-atmosphere signal dominates for both metrics. The S/N metric shows a remarkably lower observation--noise $r$-correlation than the Welch's t-test. However, $r$ increases faster with \textrm{NSF} for the S/N metric than for the Welch's t-test.
    
    \item The exo-atmospheric signal is the most noise-driven for both metrics for \textrm{NSF}\,$\in$\,$[0.5,1.0]$ ($150 \lesssim \overline{S/N_{exp}} \lesssim 300$, top-right panel in Fig.\,\ref{fig:Stat_scale_factors}A), although detections in this regime are, for the most part, significant and located at the expected $K_P$\,--\,$\text{V}_{\text{rest}}$. In other words, the exposure-qualities we aim to work with for this planet--telescope combination are also those for which the significance-metrics are the most affected by noise. This entails two main consequences: i) we are already working near the instrumental limit for this science-case (due to both smearing of the signal at longer exposures and noise-impact), and ii) exo-atmospheric detections are reliable but the computed significance-numbers need to be properly contextualized, as they may be strongly driven by noise. Interestingly, at nominal \textrm{NSF}\,$=$\,$1.0$, we observe a weak dependence for the Welch's t-test. 

    \item For lower-quality data, with $1.2$\,$<$\,\textrm{NSF}\,$<$\,$1.5$ ($100$\,$\lesssim$\,$\overline{S/N_{exp}}$\,$\lesssim$\,$125$, bottom-left panel in Fig.\,\ref{fig:Stat_scale_factors}A), the maximum-S/N signals drop below S/N\,$=$\,$4$ (Welch's $\sigma$-value\,$\sim$\,$7$) and there is a higher non-detection probability.
    Two sub-populations of nights emerge. One set in which the detection is still possible due to favoring observation--noise interplay (i.e. with high $r$), and another set for which non-detections occur and the observation--noise correlation is broken (i.e. flat distribution of points). The co-existence of these two populations yields a zero or even negative $r$-correlation. 
    
    \item For higher scaling factors ($\overline{S/N_{exp}}$\,$\lesssim$\,$100$, bottom-right panel in Fig.\,\ref{fig:Stat_scale_factors}A), the low data-quality prevents detecting exo-atmospheric signals. The CCF maps become noise-dominated, increasing $r$ until a plateau is reached for \textrm{NSF}\,$\sim$\,$4.0$. 
    The $r$-values in this region mark the basal correlation between noise-dominated data and their respective $\epsilon_N \left( \lambda,t \right)$, which is higher for Welch's t-tests.
\end{enumerate}

\subsection{False positives and bias from PCA optimization}
\label{sect:bias_opt}

The Earth's telluric contribution and the stellar spectral features are fundamentally wavelength-dependent, with some spectral regions being considerably more affected than others. This has led to the development of different methodologies to optimize the performance of PCA-based algorithms like {\tt SYSREM} so that the number of principal components subtracted (also refereed to as ``{\tt SYSREM} passes'') varies according to the level of contamination in each spectral order of the observed dataset \citep[e.g.][]{birkby2017discovery, cabot2019robustness, sanchez2019water, sanchez2022searching, Cheverall2023, holmberg2022first}. These techniques typically involve injecting signals of varying strengths relative to the expected exo-atmospheric absorption, followed by optimizing the order-wise injection recovery. The number of {\tt SYSREM} passes for each spectral order is then determined by evaluating the goodness of the injection recovery following either one of two criteria. Namely, maximizing the recovery of the injected CCF signal, or rather maximizing the difference between CCFs with and without injection. However, these procedures have been shown to inflate or even create spurious signals beyond a S/N of $\lesssim$\,$4.5$), particularly when the test injection is performed at the expected $K_P$\,--\,$\text{v}_{\text{rest}}$ of the exoplanet. This suggests that at least some optimization methods are unrobust \citep{cabot2019robustness}.

\begin{figure}[tbh!]
    \centering
    \includegraphics[width=1\columnwidth]{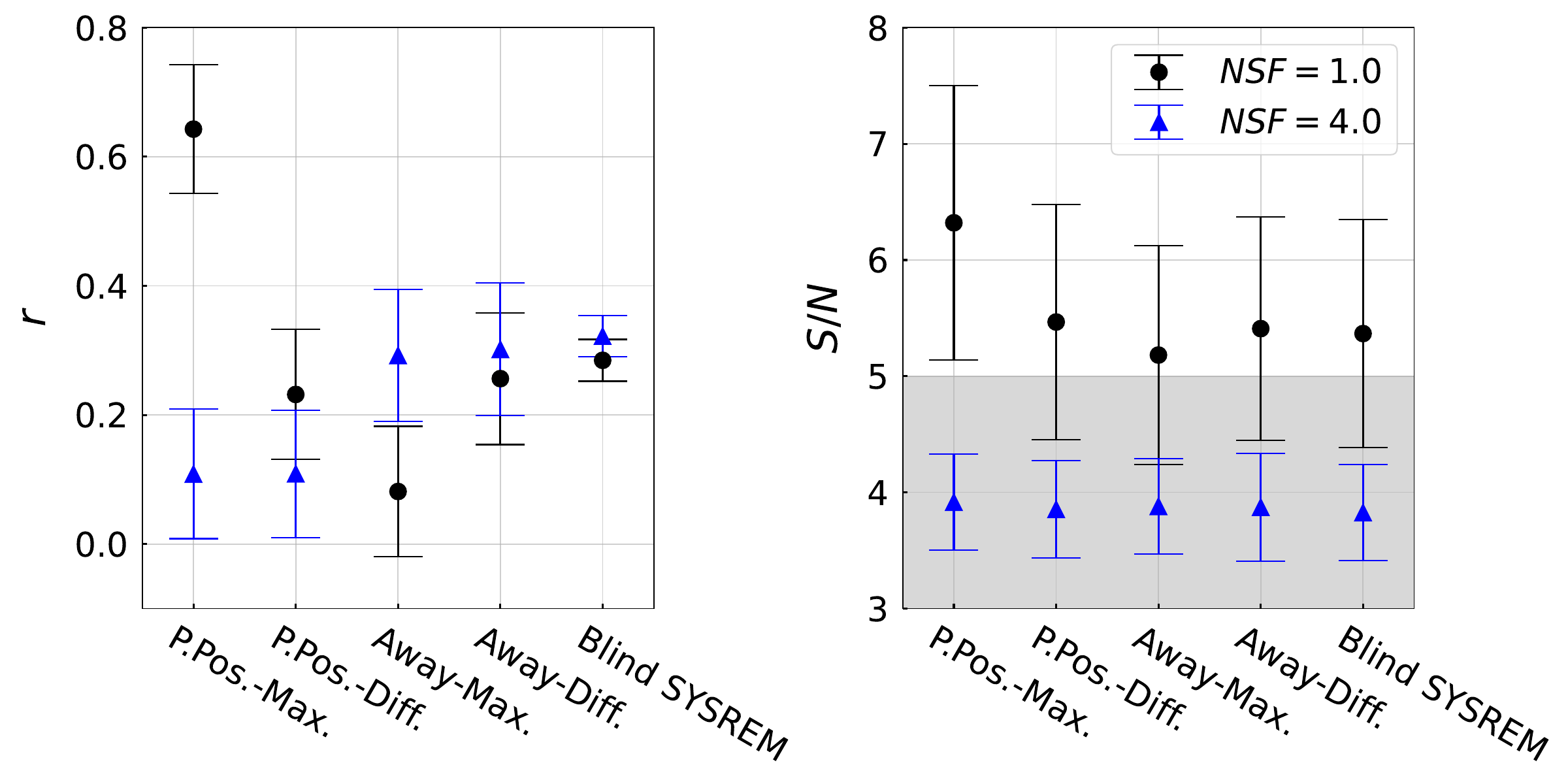}
    \caption{{Data--noise correlation for various data qualities (\textrm{NSF}\,$=$\,$1.0$, black; \textrm{NSF}\,$=$\,$4.0$, blue) and optimization methods for order-wise {\tt SYSREM} corrections. The left panel displays Pearson correlation coefficients ($r$) from $300$ simulated observations, while the right panel shows the mean S/N significance across realizations, with the gray shaded region marking S/N values below the common detection threshold of $5$. The optimization involves injecting a model H$_2$O signal at the expected planet velocities (labeled ``Planet pos.'', $K_P$\,$=$\,$152$\,\kms\ and $\text{v}_{\text{rest}}$\,$=$\,$0$\,\kms) and away from it (``Away''; at $K_P$\,$=$\,$80$\,\kms\ and $\text{V}_{\text{rest}}$\,$=$\,$80$\,\kms). Signal recovery is optimized order-by-order either by maximizing the injected CCF peak (``Max.'') or by maximizing the difference between CCFs with and without injection at the injection location (``Diff.''). Both panels also show the results from the “Blind {\tt SYSREM}” approach, applying $5$ {\tt SYSREM} passes to all spectral orders.}
    }
    \label{fig:stat_opt}
\end{figure}

The origin of inflation and of these spurious signals is challenging to pinpoint and, in the real datasets studied in the literature, may be heavily influenced by residual stellar spectral features or telluric absorption not fully removed by preparing pipelines. However, inflation may also partly come from an undesired enhancing of signal-like noise by the order-wise optimization, as we show in Sect.\,\ref{sect:source_of_var}. Utilizing the methods discussed in the previous section, we investigated the influence of noise in detection significances (i.e. $r$-correlation) when optimizing {\tt SYSREM} order by order maximizing either the recovery of an injected signal or the CCF difference with and without injection (see Fig.\,\ref{fig:stat_opt}). We performed this test for two scenarios: injecting at the true $K_P$\,--\,$\text{V}_{\text{rest}}$, and away from it. The injection is performed at three times the expected strength of absorption, so it is better identified by the optimization algorithm. At the nominal uncertainties, we found that maximizing the strength of the injected CCF signal at the exoplanet's velocities yielded the highest detection-S/N. However, this combination (``Planet pos.''\,--\,``Max.'') also shows the highest $r$-correlation with noise. It is thus likely that the higher significances are not produced by a better disentangling of the exo-atmospheric signal, but rather by the significantly higher correlation with the noise, making it the most unreliable approach. This is in line with earlier studies performed on a real dataset \citep{Cheverall2023}. Other approaches are comparably robust as per their $r$, and the S/N of the resulting CCF peaks is very similar for all of them. Additional differences may be revealed for datasets with higher telluric variability than what we included, which could be explored in future studies. Regarding noise-dominated data (\textrm{NSF}\,$=$\,$4.0$), no significant signals were detected by any method and their corresponding data--noise correlation is similar. This suggests that there are no significant advantages obtained from optimizing order-wise detrending in poor-quality datasets.

\section{Conclusion}
As exoplanet science progresses, the need for robust and reliable analysis techniques and signal-interpretations becomes increasingly critical. Evaluating and comparing different methodologies provides valuable insights into the strengths and limitations of each approach for disentangling the true nature of these distant worlds. In this work, we have successfully developed and validated {\tt EXoPLORE}, a comprehensive and customizable simulator for high-resolution exo-atmosphere observations. Through the analysis of \planeto, a well-studied hot Jupiter, we have demonstrated the robustness of our framework by reproducing previously-reported H$_2$O signals with comparable significances. Our \emph{in-silico} observations further confirmed the reliability of our approach, showing consistent signal-recoveries within 1-$\sigma$ of real-data analyses.

Simulators of high-resolution observations, such as {\tt EXoPLORE}, can be exploited to extract further insights from the data, and also from our analysis techniques. Here, we have shown that cross-correlation results depend heavily on methodological choices, including the CCF velocity steps and intervals, and the significance metrics employed. Using a CCF step smaller than the instrumental resolution causes oversampling, inflating Welch’s $\sigma$-values relative to S/N significances, while narrow velocity intervals inflate S/N by artificially decreasing the computed standard deviation of the CCF noise. These discrepancies highlight the arbitrary nature of some methodological choices and detection thresholds. It is crucial to provide detailed methodological descriptions so as to properly contextualize significances, ensure reproducibility, and facilitate comparative analyses in exo-atmospheric studies. This will be of the utmost importance to properly assess compositional and thermal variability across multiple observations with future high-quality data (e.g. with the Extremely Large Telescope).

Bayesian retrievals represent a significant advantage for in-depth characterization of exo-atmospheres, even in challenging datasets where traditional CCF methods cannot provide robust signals. Simulated datasets show that the impact of noise in signal-recoverability differs between Bayesian retrievals and CCF. In our analyses, both strong and undetectable H$_2$O signals, by S/N standards, were commonly revealed with Bayesian retrievals, hence constraining key parameters. Although different noise realizations inevitably bias retrievals as well, our findings indicate that retrieval results are more resilient, which opens a window for re-evaluating datasets previously considered inconclusive. Retrievals not only detect compounds in exo-atmospheres, but are fundamentally used to constrain physico-chemical parameters. Therefore, any reported parameter estimates should be accompanied by validation demonstrating that the retrieval framework does not introduce biases. For instance, this can be achieved by showing that the framework produces accurate and unbiased results when applied to noiseless simulated datasets.

Understanding the impact of observational noise on significance-metrics is key for accurately assessing the robustness of exo-atmospheric signals. For CCF, the measured significance depends heavily on how the spectral-noise correlates with the template used, since this interference can either amplify or suppress real signals. For weak signals (S/N\,$\lesssim$\,$4$), the challenge lies in determining in what manner noise might be interfering with a real-signal, making it difficult to assess robustness. However, finding the maximum-significance signal at the expected exoplanet's $K_P$\,--\,$\text{V}_{\text{rest}}$ can serve as a strong, albeit not infallible, indicator of veracity. While both the S/N and Welch's t-test metrics experience similar interference from noise, we find the latter is generally more robust for high exposure-quality.

Order-wise optimization of PCA-like algorithms remains challenging and can inadvertently bias the results. Our investigation using PCA-based algorithms like {\tt SYSREM} confirms the introduction of biases when optimizing the removal of the Earth's and stellar contributions. Specifically, injecting a signal at the exoplanet's $K_P$\,--\,$\text{V}_{\text{rest}}$ and maximizing the order-wise injected-CCF recovery yields the strongest biases at our nominal data-quality. Thus, higher CCF's S/N obtained with this method are likely due to preferential selection of noise.
However, order-wise optimization of the telluric correction by using an injected-signal away from the exoplanet velocities appears to be safer, but we do not observe a significant enhancement of the signal with respect to the blind approach.
For very noisy datasets, all of the tested criteria fail to recover signals. However, optimization-performance remains to be tested for simulated datasets with high telluric variability.

\section*{Acknowledgments}
We thank Siddhart Gandhi, Aurora Kesseli, Paul Molli\`ere, Doriann Blain,  Enric Pall\'e, 
and Ignas Snellen for helpful discussions. 
A.S.L. acknowledges financial support from the Severo Ochoa grant CEX2021-001131-S funded by MCIN/AEI/ 10.13039/501100011033. A.P.M. is financially supported by the ``Ram\'on y Cajal'' program of the Spanish Ministry of Science and Innovation (grant RYC2021-031241-I). A.P.M. acknowledges the Spanish Ministry and Agencia Estatal de investigación (AEI) through
Project of I+D+i (PID2020-113681GB-I00), financed by MICIN/AEI/10.13039/501100011033 and FEDER ``A way to make Europe'', and the \emph{Consejer\'ia de Conocimiento, Investigaci\'on, Universidad, Junta de Andaluc\'ia} and European
Regional Development Fund (P20-00173) for financial support. This research made use of the Spanish Virtual Observatory (http://svo.cab.inta-csic.es) supported by MINECO/FEDER through grant AyA2017-84089.7.

\bibliographystyle{plainnat} 
\bibliography{mybib}        

\begin{thebibliography}{75}
\providecommand{\natexlab}[1]{#1}
\providecommand{\url}[1]{\texttt{#1}}
\expandafter\ifx\csname urlstyle\endcsname\relax
  \providecommand{\doi}[1]{doi: #1}\else
  \providecommand{\doi}{doi: \begingroup \urlstyle{rm}\Url}\fi

\bibitem[Alonso-Floriano et~al.(2019)Alonso-Floriano, S{\'a}nchez-L{\'o}pez, Snellen, L{\'o}pez-Puertas, Nagel, Amado, Bauer, Caballero, Czesla, Nortmann, et~al.]{alonso2019multiple}
FJ~Alonso-Floriano, A~S{\'a}nchez-L{\'o}pez, IAG Snellen, Manuel L{\'o}pez-Puertas, E~Nagel, Pedro~J Amado, FF~Bauer, JA~Caballero, S~Czesla, L~Nortmann, et~al.
\newblock Multiple water band detections in the carmenes near-infrared transmission spectrum of hd 189733 b.
\newblock \emph{Astronomy \& Astrophysics}, 621:\penalty0 A74, 2019.

\bibitem[Barstow et~al.(2016)Barstow, Aigrain, Irwin, and Sing]{barstow2016consistent}
Joanna~K Barstow, Suzanne Aigrain, Patrick~GJ Irwin, and David~K Sing.
\newblock A consistent retrieval analysis of 10 hot jupiters observed in transmission.
\newblock \emph{The Astrophysical Journal}, 834\penalty0 (1):\penalty0 50, 2016.

\bibitem[Benneke et~al.(2024)Benneke, Roy, Coulombe, Radica, Piaulet, Ahrer, Pierrehumbert, Krissansen-Totton, Schlichting, Hu, et~al.]{benneke2024jwst}
Bj{\"o}rn Benneke, Pierre-Alexis Roy, Louis-Philippe Coulombe, Michael Radica, Caroline Piaulet, Eva-Maria Ahrer, Raymond Pierrehumbert, Joshua Krissansen-Totton, Hilke~E Schlichting, Renyu Hu, et~al.
\newblock Jwst reveals ch $ \_4 $, co $ \_2 $, and h $ \_2 $ o in a metal-rich miscible atmosphere on a two-earth-radius exoplanet.
\newblock \emph{arXiv preprint arXiv:2403.03325}, 2024.

\bibitem[Birkby(2018)]{birkby2018exoplanet}
Jayne~L Birkby.
\newblock Exoplanet atmospheres at high spectral resolution.
\newblock \emph{arXiv preprint arXiv:1806.04617}, 2018.

\bibitem[Birkby et~al.(2013)Birkby, De~Kok, Brogi, de~Mooij, Schwarz, Albrecht, and Snellen]{birkby2013detection}
JL~Birkby, RJ~De~Kok, M~Brogi, EJW de~Mooij, H~Schwarz, S~Albrecht, and IAG Snellen.
\newblock Detection of water absorption in the day side atmosphere of hd 189733 b using ground-based high-resolution spectroscopy at 3.2 $\mu$m.
\newblock \emph{Monthly Notices of the Royal Astronomical Society: Letters}, 436\penalty0 (1):\penalty0 L35--L39, 2013.

\bibitem[Birkby et~al.(2017)Birkby, de~Kok, Brogi, Schwarz, and Snellen]{birkby2017discovery}
JL~Birkby, RJ~de~Kok, M~Brogi, H~Schwarz, and IAG Snellen.
\newblock Discovery of water at high spectral resolution in the atmosphere of 51 peg b.
\newblock \emph{The Astronomical Journal}, 153\penalty0 (3):\penalty0 138, 2017.

\bibitem[Blain et~al.(2024{\natexlab{a}})Blain, Landman, Molli{\`e}re, and Dittmann]{blain2024four}
D~Blain, R~Landman, P~Molli{\`e}re, and J~Dittmann.
\newblock Four hd 209458 b transits through crires+: Detection of h2o and non-detections of c2h2, ch4, and hcn.
\newblock \emph{Astronomy \& Astrophysics}, 690:\penalty0 A63, 2024{\natexlab{a}}.

\bibitem[Blain et~al.(2024{\natexlab{b}})Blain, S\'anchez-L\'opez, and Molli\`ere]{blain2023retrieval}
Doriann Blain, Alejandro S\'anchez-L\'opez, and Paul Molli\`ere.
\newblock A retrieval framework applied to the high resolution transmission spectrum of hd~189733~b.
\newblock \emph{The Astronomical Journal}, X\penalty0 (X):\penalty0 X, 2024{\natexlab{b}}.

\bibitem[Boucher et~al.(2021)Boucher, Darveau-Bernier, Pelletier, Lafreni{\`e}re, Artigau, Cook, Allart, Radica, Doyon, Benneke, et~al.]{boucher2021characterizing}
Anne Boucher, Antoine Darveau-Bernier, Stefan Pelletier, David Lafreni{\`e}re, {\'E}tienne Artigau, Neil~J Cook, Romain Allart, Michael Radica, Ren{\'e} Doyon, Bj{\"o}rn Benneke, et~al.
\newblock Characterizing exoplanetary atmospheres at high resolution with spirou: detection of water on hd 189733 b.
\newblock \emph{The Astronomical Journal}, 162\penalty0 (6):\penalty0 233, 2021.

\bibitem[Brogi et~al.(2016)Brogi, De~Kok, Albrecht, Snellen, Birkby, and Schwarz]{brogi2016rotation}
M~Brogi, RJ~De~Kok, S~Albrecht, IAG Snellen, JL~Birkby, and H~Schwarz.
\newblock Rotation and winds of exoplanet hd 189733 b measured with high-dispersion transmission spectroscopy.
\newblock \emph{The Astrophysical Journal}, 817\penalty0 (2):\penalty0 106, 2016.

\bibitem[Brogi and Line(2019)]{Brogi2019}
Matteo Brogi and Michael~R. Line.
\newblock Retrieving temperatures and abundances of exoplanet atmospheres with high-resolution cross-correlation spectroscopy.
\newblock \emph{The Astronomical Journal}, 157\penalty0 (3):\penalty0 114, 02 2019.
\newblock \doi{10.3847/1538-3881/aaffd3}.
\newblock URL \url{https://doi.org/10.3847/1538-3881/aaffd3}.

\bibitem[Brogi et~al.(2012)Brogi, Snellen, De~Kok, Albrecht, Birkby, and de~Mooij]{brogi2012signature}
Matteo Brogi, Ignas~AG Snellen, Remco~J De~Kok, Simon Albrecht, Jayne Birkby, and Ernst~JW de~Mooij.
\newblock The signature of orbital motion from the dayside of the planet $\tau$ bo{\"o}tis b.
\newblock \emph{Nature}, 486\penalty0 (7404):\penalty0 502--504, 2012.

\bibitem[Brogi et~al.(2018)Brogi, Giacobbe, Guilluy, de~Kok, Sozzetti, Mancini, and Bonomo]{brogi2018exoplanet}
Matteo Brogi, P~Giacobbe, G~Guilluy, RJ~de~Kok, Alessandro Sozzetti, L~Mancini, and ALDO~STEFANO Bonomo.
\newblock Exoplanet atmospheres with giano-i. water in the transmission spectrum of hd 189 733 b.
\newblock \emph{Astronomy \& Astrophysics}, 615:\penalty0 A16, 2018.

\bibitem[Caballero et~al.(2016)Caballero, Gu{\`a}rdia, del Fresno, Zechmeister, de~Juan, Alonso-Floriano, Amado, Colom{\'e}, Cort{\'e}s-Contreras, Garc{\'\i}a-Piquer, et~al.]{caballero2016carmenes}
JA~Caballero, J~Gu{\`a}rdia, M~L{\'o}pez del Fresno, M~Zechmeister, E~de~Juan, FJ~Alonso-Floriano, PJ~Amado, J~Colom{\'e}, M~Cort{\'e}s-Contreras, {\'A}~Garc{\'\i}a-Piquer, et~al.
\newblock Carmenes: data flow.
\newblock In \emph{Observatory operations: strategies, processes, and systems VI}, volume 9910, page 99100E. International Society for Optics and Photonics, 2016.

\bibitem[Cabot et~al.(2019)Cabot, Madhusudhan, Hawker, and Gandhi]{cabot2019robustness}
Samuel~HC Cabot, Nikku Madhusudhan, George~A Hawker, and Siddharth Gandhi.
\newblock On the robustness of analysis techniques for molecular detections using high-resolution exoplanet spectroscopy.
\newblock \emph{Monthly Notices of the Royal Astronomical Society}, 482\penalty0 (4):\penalty0 4422--4436, 2019.

\bibitem[Charbonneau et~al.(2002)Charbonneau, Brown, Noyes, and Gilliland]{charbonneau2002detection}
David Charbonneau, Timothy~M Brown, Robert~W Noyes, and Ronald~L Gilliland.
\newblock Detection of an extrasolar planet atmosphere.
\newblock \emph{The Astrophysical Journal}, 568\penalty0 (1):\penalty0 377, 2002.

\bibitem[Cheverall et~al.(2023)Cheverall, Madhusudhan, and Holmberg]{Cheverall2023}
Connor~J Cheverall, Nikku Madhusudhan, and Måns Holmberg.
\newblock {Robustness measures for molecular detections using high-resolution transmission spectroscopy of exoplanets}.
\newblock \emph{Monthly Notices of the Royal Astronomical Society}, 522\penalty0 (1):\penalty0 661--677, 04 2023.
\newblock ISSN 0035-8711.
\newblock \doi{10.1093/mnras/stad648}.
\newblock URL \url{https://doi.org/10.1093/mnras/stad648}.

\bibitem[Chubb and Min(2022)]{chubb2022exoplanet}
Katy~L Chubb and Michiel Min.
\newblock Exoplanet atmosphere retrievals in 3d using phase curve data with arcis: application to wasp-43b.
\newblock \emph{Astronomy \& Astrophysics}, 665:\penalty0 A2, 2022.

\bibitem[Cont et~al.(2021)Cont, Yan, Reiners, Casasayas-Barris, Molli{\`e}re, Pall{\'e}, Henning, Nortmann, Stangret, Czesla, et~al.]{cont2021detection}
D~Cont, F~Yan, A~Reiners, N~Casasayas-Barris, P~Molli{\`e}re, E~Pall{\'e}, Th~Henning, L~Nortmann, M~Stangret, S~Czesla, et~al.
\newblock Detection of fe and evidence for tio in the dayside emission spectrum of wasp-33b.
\newblock \emph{Astronomy \& Astrophysics}, 651:\penalty0 A33, 2021.

\bibitem[Cont et~al.(2022{\natexlab{a}})Cont, Yan, Reiners, Nortmann, Molaverdikhani, Pall{\'e}, Stangret, Henning, Ribas, Quirrenbach, et~al.]{cont2022silicon}
D~Cont, F~Yan, A~Reiners, L~Nortmann, K~Molaverdikhani, E~Pall{\'e}, M~Stangret, Th~Henning, I~Ribas, A~Quirrenbach, et~al.
\newblock Silicon in the dayside atmospheres of two ultra-hot jupiters.
\newblock \emph{Astronomy \& Astrophysics}, 657:\penalty0 L2, 2022{\natexlab{a}}.

\bibitem[Cont et~al.(2022{\natexlab{b}})Cont, Yan, Reiners, Nortmann, Molaverdikhani, Pall{\'e}, Henning, Ribas, Quirrenbach, Caballero, et~al.]{cont2022atmospheric}
D~Cont, F~Yan, Ansgar Reiners, L~Nortmann, K~Molaverdikhani, Enric Pall{\'e}, Th~Henning, Ignasi Ribas, Andreas Quirrenbach, JA~Caballero, et~al.
\newblock Atmospheric characterization of the ultra-hot jupiter wasp-33b-detection of ti and v emission lines and retrieval of a broadened line profile.
\newblock \emph{Astronomy \& Astrophysics}, 668:\penalty0 A53, 2022{\natexlab{b}}.

\bibitem[Danielski et~al.(2014)Danielski, Deroo, Waldmann, Hollis, Tinetti, and Swain]{danielski20140}
C~Danielski, P~Deroo, IP~Waldmann, MDJ Hollis, G~Tinetti, and MR~Swain.
\newblock 0.94--2.42 $\mu$m ground-based transmission spectra of the hot jupiter hd-189733b.
\newblock \emph{The Astrophysical Journal}, 785\penalty0 (1):\penalty0 35, 2014.

\bibitem[Dash et~al.(2024)Dash, Brogi, Gandhi, Lafarga, Meech, Bello-Arufe, and Wheatley]{dash2024constraints}
Spandan Dash, Matteo Brogi, Siddharth Gandhi, Marina Lafarga, Annabella Meech, Aaron Bello-Arufe, and Peter~J Wheatley.
\newblock Constraints on atmospheric water abundance and cloud deck pressure in the warm neptune gj 3470 b via carmenes transmission spectroscopy.
\newblock \emph{Monthly Notices of the Royal Astronomical Society}, 530\penalty0 (3):\penalty0 3100--3116, 2024.

\bibitem[Ehrenreich et~al.(2020)Ehrenreich, Lovis, Allart, Osorio, Pepe, Cristiani, Rebolo, Santos, Borsa, Demangeon, et~al.]{ehrenreich2020nightside}
David Ehrenreich, Christophe Lovis, Romain Allart, Mar{\'\i}a Rosa~Zapatero Osorio, Francesco Pepe, Stefano Cristiani, Rafael Rebolo, Nuno~C Santos, Francesco Borsa, Olivier Demangeon, et~al.
\newblock Nightside condensation of iron in an ultrahot giant exoplanet.
\newblock \emph{Nature}, 580\penalty0 (7805):\penalty0 597--601, 2020.

\bibitem[Finnerty et~al.(2024)Finnerty, Xuan, Xin, Liberman, Schofield, Fitzgerald, Agrawal, Baker, Bartos, Blake, et~al.]{finnerty2024atmospheric}
Luke Finnerty, Jerry~W Xuan, Yinzi Xin, Joshua Liberman, Tobias Schofield, Michael~P Fitzgerald, Shubh Agrawal, Ashley Baker, Randall Bartos, Geoffrey~A Blake, et~al.
\newblock Atmospheric metallicity and c/o of hd 189733 b from high-resolution spectroscopy.
\newblock \emph{The Astronomical Journal}, 167\penalty0 (1):\penalty0 43, 2024.

\bibitem[Gibson et~al.(2020)Gibson, Merritt, Nugroho, Cubillos, de~Mooij, Mikal-Evans, Fossati, Lothringer, Nikolov, Sing, et~al.]{gibson2020detection}
Neale~P Gibson, Stephanie Merritt, Stevanus~K Nugroho, Patricio~E Cubillos, Ernst~JW de~Mooij, Thomas Mikal-Evans, Luca Fossati, Joshua Lothringer, Nikolay Nikolov, David~K Sing, et~al.
\newblock Detection of fe i in the atmosphere of the ultra-hot jupiter wasp-121b, and a new likelihood-based approach for doppler-resolved spectroscopy.
\newblock \emph{Monthly Notices of the Royal Astronomical Society}, 493\penalty0 (2):\penalty0 2215--2228, 2020.

\bibitem[Gibson et~al.(2022)Gibson, Nugroho, Lothringer, Maguire, and Sing]{gibson2022relative}
Neale~P Gibson, Stevanus~K Nugroho, Joshua Lothringer, Cathal Maguire, and David~K Sing.
\newblock Relative abundance constraints from high-resolution optical transmission spectroscopy of wasp-121b, and a fast model-filtering technique for accelerating retrievals.
\newblock \emph{Monthly Notices of the Royal Astronomical Society}, 512\penalty0 (3):\penalty0 4618--4638, 2022.

\bibitem[Guillot(2010)]{guillot2010radiative}
Tristan Guillot.
\newblock On the radiative equilibrium of irradiated planetary atmospheres.
\newblock \emph{Astronomy \& Astrophysics}, 520:\penalty0 A27, 2010.

\bibitem[Hawker et~al.(2018)Hawker, Madhusudhan, Cabot, and Gandhi]{hawker2018evidence}
George~A Hawker, Nikku Madhusudhan, Samuel~HC Cabot, and Siddharth Gandhi.
\newblock Evidence for multiple molecular species in the hot jupiter hd 209458b.
\newblock \emph{The Astrophysical Journal Letters}, 863\penalty0 (1):\penalty0 L11, 2018.

\bibitem[Hoeijmakers et~al.(2018)Hoeijmakers, Ehrenreich, Heng, Kitzmann, Grimm, Allart, Deitrick, Wyttenbach, Oreshenko, Pino, et~al.]{hoeijmakers2018atomic}
H~Jens Hoeijmakers, David Ehrenreich, Kevin Heng, Daniel Kitzmann, Simon~L Grimm, Romain Allart, Russell Deitrick, Aur{\'e}lien Wyttenbach, Maria Oreshenko, Lorenzo Pino, et~al.
\newblock Atomic iron and titanium in the atmosphere of the exoplanet kelt-9b.
\newblock \emph{Nature}, 560\penalty0 (7719):\penalty0 453--455, 2018.

\bibitem[Hoeijmakers et~al.(2020)Hoeijmakers, Cabot, Zhao, Buchhave, Tronsgaard, Davis, Kitzmann, Grimm, Cegla, Bourrier, et~al.]{hoeijmakers2020high}
H~Jens Hoeijmakers, Samuel~HC Cabot, Lily Zhao, Lars~A Buchhave, Ren{\'e} Tronsgaard, Allen~B Davis, Daniel Kitzmann, Simon~L Grimm, Heather~M Cegla, Vincent Bourrier, et~al.
\newblock High-resolution transmission spectroscopy of mascara-2 b with expres.
\newblock \emph{Astronomy \& Astrophysics}, 641:\penalty0 A120, 2020.

\bibitem[Hoeijmakers et~al.(2019)Hoeijmakers, Ehrenreich, Kitzmann, Allart, Grimm, Seidel, Wyttenbach, Pino, Nielsen, Fisher, et~al.]{hoeijmakers2019spectral}
Herman~Jens Hoeijmakers, D~Ehrenreich, Daniel Kitzmann, R~Allart, SL~Grimm, JV~Seidel, A~Wyttenbach, Lorenzo Pino, LD~Nielsen, C~Fisher, et~al.
\newblock A spectral survey of an ultra-hot jupiter-detection of metals in the transmission spectrum of kelt-9 b.
\newblock \emph{Astronomy \& Astrophysics}, 627:\penalty0 A165, 2019.

\bibitem[Holmberg and Madhusudhan(2022)]{holmberg2022first}
M{\aa}ns Holmberg and Nikku Madhusudhan.
\newblock A first look at crires+: Performance assessment and exoplanet spectroscopy.
\newblock \emph{The Astronomical Journal}, 164\penalty0 (3):\penalty0 79, 2022.

\bibitem[Holmberg and Madhusudhan(2024)]{holmberg2024possible}
M{\aa}ns Holmberg and Nikku Madhusudhan.
\newblock Possible hycean conditions in the sub-neptune toi-270 d.
\newblock \emph{Astronomy \& Astrophysics}, 683:\penalty0 L2, 2024.

\bibitem[Jones et~al.(2013)Jones, Noll, Kausch, Szyszka, and Kimeswenger]{jones2013advanced}
Amy Jones, Stefan Noll, Wolfgang Kausch, Cezary Szyszka, and Stefan Kimeswenger.
\newblock An advanced scattered moonlight model for cerro paranal.
\newblock \emph{Astronomy \& Astrophysics}, 560:\penalty0 A91, 2013.

\bibitem[Kesseli and Snellen(2021)]{kesseli2021confirmation}
Aurora~Y Kesseli and IAG Snellen.
\newblock Confirmation of asymmetric iron absorption in wasp-76b with harps.
\newblock \emph{The Astrophysical Journal Letters}, 908\penalty0 (1):\penalty0 L17, 2021.

\bibitem[Kesseli et~al.(2020)Kesseli, Snellen, Alonso-Floriano, Molli{\`e}re, and Serindag]{kesseli2020search}
Aurora~Y Kesseli, IAG Snellen, FJ~Alonso-Floriano, P~Molli{\`e}re, and DB~Serindag.
\newblock A search for feh in hot-jupiter atmospheres with high-dispersion spectroscopy.
\newblock \emph{The Astronomical Journal}, 160\penalty0 (5):\penalty0 228, 2020.

\bibitem[Kesseli et~al.(2022)Kesseli, Snellen, Casasayas-Barris, Molli{\`e}re, and S{\'a}nchez-L{\'o}pez]{kesseli2022atomic}
Aurora~Y Kesseli, IAG Snellen, N~Casasayas-Barris, P~Molli{\`e}re, and A~S{\'a}nchez-L{\'o}pez.
\newblock An atomic spectral survey of wasp-76b: Resolving chemical gradients and asymmetries.
\newblock \emph{The Astronomical Journal}, 163\penalty0 (3):\penalty0 107, 2022.

\bibitem[Klein et~al.(2024)Klein, Debras, Donati, Hood, Moutou, Carmona, Ould-Elkhim, B{\'e}zard, Charnay, Fouqu{\'e}, et~al.]{klein2024atmospherix}
Baptiste Klein, Florian Debras, Jean-Fran{\c{c}}ois Donati, Thea Hood, Claire Moutou, Andres Carmona, Merwan Ould-Elkhim, Bruno B{\'e}zard, Benjamin Charnay, Pascal Fouqu{\'e}, et~al.
\newblock Atmospherix: I-an open source high-resolution transmission spectroscopy pipeline for exoplanets atmospheres with spirou.
\newblock \emph{Monthly Notices of the Royal Astronomical Society}, 527\penalty0 (1):\penalty0 544--565, 2024.

\bibitem[{Kreidberg}(2015)]{kreidberg2015batman2}
Laura {Kreidberg}.
\newblock {batman: BAsic Transit Model cAlculatioN in Python}.
\newblock \emph{Publications of the Astronomical Society of the Pacific (PASP)}, 127\penalty0 (957):\penalty0 1161, November 2015.
\newblock \doi{10.1086/683602}.

\bibitem[Landman et~al.(2021)Landman, S\'anchez-L\'opez, Molli\`ere, Kesseli, Louca, and Snellen]{landman2021}
Rico Landman, Alejandro S\'anchez-L\'opez, Paul Molli\`ere, Aurora~Y Kesseli, AJ~Louca, and IAG Snellen.
\newblock Detection of oh in the ultra-hot jupiter wasp-76b.
\newblock \emph{Astronomy \& Astrophysics}, 656:\penalty0 A119, 2021.

\bibitem[Lesjak et~al.(2023)Lesjak, Nortmann, Yan, Cont, Reiners, Piskunov, Hatzes, Boldt-Christmas, Czesla, Heiter, et~al.]{lesjak2023retrieval}
F~Lesjak, L~Nortmann, F~Yan, D~Cont, A~Reiners, Nikolai Piskunov, A~Hatzes, Linn Boldt-Christmas, S~Czesla, Ulrike Heiter, et~al.
\newblock Retrieval of the dayside atmosphere of wasp-43b with crires+.
\newblock \emph{Astronomy \& Astrophysics}, 678:\penalty0 A23, 2023.

\bibitem[Madhusudhan(2019)]{madhusudhan2019exoplanetary}
Nikku Madhusudhan.
\newblock Exoplanetary atmospheres: key insights, challenges, and prospects.
\newblock \emph{Annual Review of Astronomy and Astrophysics}, 57\penalty0 (1):\penalty0 617--663, 2019.

\bibitem[Madhusudhan et~al.(2014)Madhusudhan, Crouzet, McCullough, Deming, and Hedges]{madhusudhan2014h2o}
Nikku Madhusudhan, Nicolas Crouzet, Peter~R McCullough, Drake Deming, and Christina Hedges.
\newblock H2o abundances in the atmospheres of three hot jupiters.
\newblock \emph{The Astrophysical Journal Letters}, 791\penalty0 (1):\penalty0 L9, 2014.

\bibitem[Maguire et~al.(2024)Maguire, Gibson, Nugroho, Fortune, Ramkumar, Gandhi, and de~Mooij]{maguire2024high}
Cathal Maguire, Neale~P Gibson, Stevanus~K Nugroho, Mark Fortune, Swaetha Ramkumar, Siddharth Gandhi, and Ernst de~Mooij.
\newblock High-resolution atmospheric retrievals of wasp-76b transmission spectroscopy with espresso: Monitoring limb asymmetries across multiple transits.
\newblock \emph{arXiv preprint arXiv:2404.10463}, 2024.

\bibitem[Mansfield et~al.(2024)Mansfield, Line, Wardenier, Brogi, Bean, Beltz, Smith, Zalesky, Batalha, Kempton, et~al.]{mansfield2024metallicity}
Megan~Weiner Mansfield, Michael~R Line, Joost~P Wardenier, Matteo Brogi, Jacob~L Bean, Hayley Beltz, Peter Smith, Joseph~A Zalesky, Natasha Batalha, Eliza M-R Kempton, et~al.
\newblock The metallicity and carbon-to-oxygen ratio of the ultrahot jupiter wasp-76b from gemini-s/igrins.
\newblock \emph{The Astronomical Journal}, 168\penalty0 (1):\penalty0 14, 2024.

\bibitem[Mazeh et~al.(2007)Mazeh, Tamuz, and Zucker]{mazeh2007transiting}
T~Mazeh, O~Tamuz, and S~Zucker.
\newblock Transiting extrasolar planets workshop.
\newblock In \emph{ASP Conf. Ser.}, volume 366, page 119, 2007.

\bibitem[McCullough et~al.(2014)McCullough, Crouzet, Deming, and Madhusudhan]{mccullough2014water}
PR~McCullough, N~Crouzet, D~Deming, and N~Madhusudhan.
\newblock Water vapor in the spectrum of the extrasolar planet hd 189733b. i. the transit.
\newblock \emph{The Astrophysical Journal}, 791\penalty0 (1):\penalty0 55, 2014.

\bibitem[Merritt et~al.(2020)Merritt, Gibson, Nugroho, de~Mooij, Hooton, Matthews, McKemmish, Mikal-Evans, Nikolov, Sing, et~al.]{merritt2020non}
Stephanie~R Merritt, Neale~P Gibson, Stevanus~K Nugroho, Ernst~JW de~Mooij, Matthew~J Hooton, Shannon~M Matthews, Laura~K McKemmish, Thomas Mikal-Evans, Nikolay Nikolov, David~K Sing, et~al.
\newblock Non-detection of tio and vo in the atmosphere of wasp-121b using high-resolution spectroscopy.
\newblock \emph{Astronomy \& Astrophysics}, 636:\penalty0 A117, 2020.

\bibitem[Molli{\`e}re et~al.(2019)Molli{\`e}re, Wardenier, van Boekel, Henning, Molaverdikhani, and Snellen]{molliere2019petitradtrans}
P~Molli{\`e}re, JP~Wardenier, R~van Boekel, Th~Henning, K~Molaverdikhani, and IAG Snellen.
\newblock petitradtrans-a python radiative transfer package for exoplanet characterization and retrieval.
\newblock \emph{Astronomy \& Astrophysics}, 627:\penalty0 A67, 2019.

\bibitem[Noll et~al.(2012)Noll, Kausch, Barden, Jones, Szyszka, Kimeswenger, and Vinther]{noll2012atmospheric}
S~Noll, W~Kausch, M~Barden, AM~Jones, C~Szyszka, S~Kimeswenger, and J~Vinther.
\newblock An atmospheric radiation model for cerro paranal-i. the optical spectral range.
\newblock \emph{Astronomy \& Astrophysics}, 543:\penalty0 A92, 2012.

\bibitem[Nortmann et~al.(2024)Nortmann, Lesjak, Yan, Cont, Czesla, Lavail, Rains, Nagel, Boldt-Christmas, Hatzes, et~al.]{nortmann2024crires}
L~Nortmann, F~Lesjak, F~Yan, D~Cont, S~Czesla, A~Lavail, AD~Rains, E~Nagel, L~Boldt-Christmas, A~Hatzes, et~al.
\newblock Crires $\hat{+}$ transmission spectroscopy of wasp-127b. detection of the resolved signatures of a supersonic equatorial jet and cool poles in a hot planet.
\newblock \emph{arXiv preprint arXiv:2404.12363}, 2024.

\bibitem[Nortmann et~al.(2018)Nortmann, Pall{\'e}, Salz, Sanz-Forcada, Nagel, Alonso-Floriano, Czesla, Yan, Chen, Snellen, et~al.]{nortmann2018ground}
Lisa Nortmann, Enric Pall{\'e}, Michael Salz, Jorge Sanz-Forcada, Evangelos Nagel, F~Javier Alonso-Floriano, Stefan Czesla, Fei Yan, Guo Chen, Ignas~AG Snellen, et~al.
\newblock Ground-based detection of an extended helium atmosphere in the saturn-mass exoplanet wasp-69b.
\newblock \emph{Science}, 362\penalty0 (6421):\penalty0 1388--1391, 2018.

\bibitem[Nugroho et~al.(2017)Nugroho, Kawahara, Masuda, Hirano, Kotani, and Tajitsu]{nugroho2017high}
Stevanus~K Nugroho, Hajime Kawahara, Kento Masuda, Teruyuki Hirano, Takayuki Kotani, and Akito Tajitsu.
\newblock High-resolution spectroscopic detection of tio and a stratosphere in the day-side of wasp-33b.
\newblock \emph{The Astronomical Journal}, 154\penalty0 (6):\penalty0 221, 2017.

\bibitem[Nugroho et~al.(2020)Nugroho, Gibson, de~Mooij, Watson, Kawahara, and Merritt]{nugroho2020searching}
Stevanus~K Nugroho, Neale~P Gibson, Ernst~JW de~Mooij, Chris~A Watson, Hajime Kawahara, and Stephanie Merritt.
\newblock Searching for thermal inversion agents in the transmission spectrum of kelt-20b/mascara-2b: detection of neutral iron and ionised calcium h\&k lines.
\newblock \emph{Monthly Notices of the Royal Astronomical Society}, 496\penalty0 (1):\penalty0 504--522, 2020.

\bibitem[Nugroho et~al.(2021)Nugroho, Kawahara, Gibson, de~Mooij, Hirano, Kotani, Kawashima, Masuda, Brogi, Birkby, et~al.]{nugroho2021first}
Stevanus~K Nugroho, Hajime Kawahara, Neale~P Gibson, Ernst~JW de~Mooij, Teruyuki Hirano, Takayuki Kotani, Yui Kawashima, Kento Masuda, Matteo Brogi, Jayne~L Birkby, et~al.
\newblock First detection of hydroxyl radical emission from an exoplanet atmosphere: High-dispersion characterization of wasp-33b using subaru/ird.
\newblock \emph{The Astrophysical Journal Letters}, 910\penalty0 (1):\penalty0 L9, 2021.

\bibitem[Pinhas et~al.(2019)Pinhas, Madhusudhan, Gandhi, and MacDonald]{pinhas2019h2o}
Arazi Pinhas, Nikku Madhusudhan, Siddharth Gandhi, and Ryan MacDonald.
\newblock H2o abundances and cloud properties in ten hot giant exoplanets.
\newblock \emph{Monthly Notices of the Royal Astronomical Society}, 482\penalty0 (2):\penalty0 1485--1498, 2019.

\bibitem[Quirrenbach et~al.(2016)Quirrenbach, Amado, Caballero, Mundt, Reiners, Ribas, Seifert, Abril, Aceituno, Alonso-Floriano, et~al.]{quirrenbach2016carmenes}
A~Quirrenbach, PJ~Amado, JA~Caballero, R~Mundt, A~Reiners, I~Ribas, W~Seifert, M~Abril, J~Aceituno, FJ~Alonso-Floriano, et~al.
\newblock Carmenes: an overview six months after first light.
\newblock In \emph{Ground-based and Airborne Instrumentation for Astronomy VI}, volume 9908, page 990812. International Society for Optics and Photonics, 2016.

\bibitem[Quirrenbach et~al.(2018)Quirrenbach, Amado, Ribas, Reiners, Caballero, Seifert, Aceituno, Azzaro, Baroch, Barrado, et~al.]{quirrenbach2018carmenes}
A~Quirrenbach, Pedro~J Amado, I~Ribas, A~Reiners, JA~Caballero, W~Seifert, J~Aceituno, M~Azzaro, D~Baroch, D~Barrado, et~al.
\newblock Carmenes: high-resolution spectra and precise radial velocities in the red and infrared.
\newblock In \emph{Ground-based and Airborne Instrumentation for Astronomy VII}, volume 10702, page 107020W. International Society for Optics and Photonics, 2018.

\bibitem[Rafi et~al.(2024)Rafi, Nugroho, Tamura, Nortmann, and S{\'a}nchez-L{\'o}pez]{rafi2024evidence}
Sayyed~A Rafi, Stevanus~K Nugroho, Motohide Tamura, Lisa Nortmann, and Alejandro S{\'a}nchez-L{\'o}pez.
\newblock Evidence of water vapor in the atmosphere of a metal-rich hot saturn with high-resolution transmission spectroscopy.
\newblock \emph{The Astronomical Journal}, 168\penalty0 (3):\penalty0 106, 2024.

\bibitem[Redfield et~al.(2008)Redfield, Endl, Cochran, and Koesterke]{redfield2008sodium}
Seth Redfield, Michael Endl, William~D Cochran, and Lars Koesterke.
\newblock Sodium absorption from the exoplanetary atmosphere of hd 189733b detected in the optical transmission spectrum.
\newblock \emph{The Astrophysical Journal}, 673\penalty0 (1):\penalty0 L87, 2008.

\bibitem[S{\'a}nchez-L{\'o}pez et~al.(2019)S{\'a}nchez-L{\'o}pez, Alonso-Floriano, L{\'o}pez-Puertas, Snellen, Funke, Nagel, Bauer, Amado, Caballero, Czesla, et~al.]{sanchez2019water}
A~S{\'a}nchez-L{\'o}pez, FJ~Alonso-Floriano, M~L{\'o}pez-Puertas, IAG Snellen, B~Funke, E~Nagel, FF~Bauer, PJ~Amado, JA~Caballero, S~Czesla, et~al.
\newblock Water vapor detection in the transmission spectra of hd 209458 b with the carmenes nir channel.
\newblock \emph{A\&A}, 630:\penalty0 A53, 2019.

\bibitem[S{\'a}nchez-L{\'o}pez et~al.(2020)S{\'a}nchez-L{\'o}pez, L{\'o}pez-Puertas, Snellen, Nagel, Bauer, Pall{\'e}, Tal-Or, Amado, Caballero, Czesla, et~al.]{sanchez2020discriminating}
A~S{\'a}nchez-L{\'o}pez, M~L{\'o}pez-Puertas, IAG Snellen, E~Nagel, FF~Bauer, E~Pall{\'e}, L~Tal-Or, PJ~Amado, JA~Caballero, S~Czesla, et~al.
\newblock Discriminating between hazy and clear hot-jupiter atmospheres with carmenes.
\newblock \emph{Astronomy \& Astrophysics}, 643:\penalty0 A24, 2020.

\bibitem[S{\'a}nchez-L{\'o}pez et~al.(2022)S{\'a}nchez-L{\'o}pez, Landman, Molli{\`e}re, Casasayas-Barris, Kesseli, and Snellen]{sanchez2022searching}
A~S{\'a}nchez-L{\'o}pez, R~Landman, P~Molli{\`e}re, N~Casasayas-Barris, AY~Kesseli, and IAG Snellen.
\newblock Searching for the origin of the ehrenreich effect in ultra-hot jupiters: Evidence for strong c/o gradients in the atmosphere of wasp-76b?
\newblock \emph{Astronomy \& Astrophysics}, 661:\penalty0 A78, 2022.

\bibitem[Snellen et~al.(2008)Snellen, Albrecht, De~Mooij, and Le~Poole]{snellen2008ground}
IAG Snellen, S~Albrecht, EJW De~Mooij, and RS~Le~Poole.
\newblock Ground-based detection of sodium in the transmission spectrum of exoplanet hd 209458b.
\newblock \emph{Astronomy \& Astrophysics}, 487\penalty0 (1):\penalty0 357--362, 2008.

\bibitem[Snellen et~al.(2010)Snellen, De~Kok, De~Mooij, and Albrecht]{snellen2010orbital}
Ignas~AG Snellen, Remco~J De~Kok, Ernst~JW De~Mooij, and Simon Albrecht.
\newblock The orbital motion, absolute mass and high-altitude winds of exoplanet hd 209458b.
\newblock \emph{Nature}, 465\penalty0 (7301):\penalty0 1049--1051, 2010.

\bibitem[Stangret et~al.(2020)Stangret, Casasayas-Barris, Pall{\'e}, Yan, S{\'a}nchez-L{\'o}pez, and L{\'o}pez-Puertas]{stangret2020detection}
M~Stangret, N~Casasayas-Barris, E~Pall{\'e}, F~Yan, A~S{\'a}nchez-L{\'o}pez, and Manuel L{\'o}pez-Puertas.
\newblock Detection of fe i and fe ii in the atmosphere of mascara-2b using a cross-correlation method.
\newblock \emph{Astronomy \& Astrophysics}, 638:\penalty0 A26, 2020.

\bibitem[Tamuz et~al.(2005)Tamuz, Mazeh, and Zucker]{tamuz2005correcting}
Omer Tamuz, Tsevi Mazeh, and Shay Zucker.
\newblock Correcting systematic effects in a large set of photometric light curves.
\newblock \emph{Monthly Notices of the Royal Astronomical Society}, 356\penalty0 (4):\penalty0 1466--1470, 2005.

\bibitem[Tsiaras et~al.(2018)Tsiaras, Waldmann, Zingales, Rocchetto, Morello, Damiano, Karpouzas, Tinetti, McKemmish, Tennyson, et~al.]{tsiaras2018population}
A~Tsiaras, IP~Waldmann, T~Zingales, M~Rocchetto, G~Morello, M~Damiano, K~Karpouzas, G~Tinetti, LK~McKemmish, J~Tennyson, et~al.
\newblock A population study of gaseous exoplanets.
\newblock \emph{The Astronomical Journal}, 155\penalty0 (4):\penalty0 156, 2018.

\bibitem[Vidal-Madjar et~al.(2003)Vidal-Madjar, Des~Etangs, D{\'e}sert, Ballester, Ferlet, H{\'e}brard, and Mayor]{vidal2003extended}
A~Vidal-Madjar, A~Lecavelier Des~Etangs, J-M D{\'e}sert, GE~Ballester, R~Ferlet, G~H{\'e}brard, and M~Mayor.
\newblock An extended upper atmosphere around the extrasolar planet hd209458b.
\newblock \emph{Nature}, 422\penalty0 (6928):\penalty0 143--146, 2003.

\bibitem[Welbanks et~al.(2019)Welbanks, Madhusudhan, Allard, Hubeny, Spiegelman, and Leininger]{welbanks2019mass}
Luis Welbanks, Nikku Madhusudhan, Nicole~F Allard, Ivan Hubeny, Fernand Spiegelman, and Thierry Leininger.
\newblock Mass--metallicity trends in transiting exoplanets from atmospheric abundances of h2o, na, and k.
\newblock \emph{The Astrophysical Journal}, 887\penalty0 (1):\penalty0 L20, 2019.

\bibitem[Welch(1947)]{welch1947generalization}
Bernard~L Welch.
\newblock The generalization of ‘student's’problem when several different population varlances are involved.
\newblock \emph{Biometrika}, 34\penalty0 (1-2):\penalty0 28--35, 1947.

\bibitem[Yan et~al.(2020)Yan, Espinoza, Molaverdikhani, Henning, Mancini, Mallonn, Rackham, Apai, Jord{\'a}n, Molli{\`e}re, et~al.]{yan2020lbt}
F~Yan, N~Espinoza, K~Molaverdikhani, Th~Henning, L~Mancini, M~Mallonn, BV~Rackham, D~Apai, A~Jord{\'a}n, P~Molli{\`e}re, et~al.
\newblock Lbt transmission spectroscopy of hat-p-12b-confirmation of a cloudy atmosphere with no significant alkali features.
\newblock \emph{Astronomy \& Astrophysics}, 642:\penalty0 A98, 2020.

\bibitem[Yan et~al.(2023)Yan, Nortmann, Reiners, Piskunov, Hatzes, Seemann, Shulyak, Lavail, Rains, Cont, et~al.]{yan2023crires}
F~Yan, L~Nortmann, A~Reiners, Nikolai Piskunov, A~Hatzes, U~Seemann, D~Shulyak, Alexis Lavail, AD~Rains, D~Cont, et~al.
\newblock Crires+ detection of co emissions lines and temperature inversions on the dayside of wasp-18b and wasp-76b.
\newblock \emph{Astronomy \& Astrophysics}, 672:\penalty0 A107, 2023.

\bibitem[Zechmeister et~al.(2014)Zechmeister, Anglada-Escud{\'e}, and Reiners]{zechmeister2014flat}
M~Zechmeister, G~Anglada-Escud{\'e}, and A~Reiners.
\newblock Flat-relative optimal extraction-a quick and efficient algorithm for stabilised spectrographs.
\newblock \emph{Astronomy \& Astrophysics}, 561:\penalty0 A59, 2014.

\end{thebibliography}

\newpage
\appendix
\section*{Supplementary Figures}
\addcontentsline{toc}{section}{Appendix} 
\renewcommand{\thefigure}{S\arabic{figure}} 
\setcounter{figure}{0}                      

\begin{figure}[htb!]
\centering
\includegraphics[angle=0, width=1.\columnwidth]{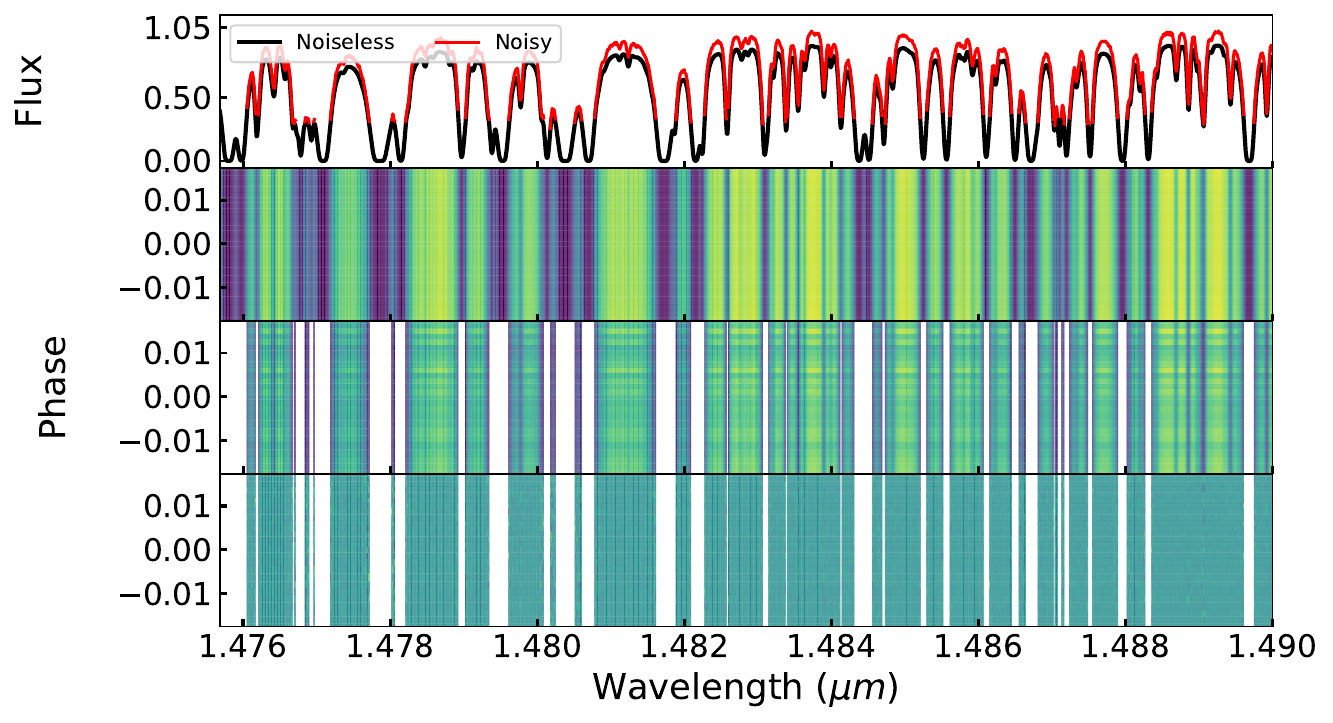}
\caption{Illustration of the simulated observations and the main analysis-steps in an arbitrary near-infrared spectral range. The first row displays the noiseless flux, which combines contributions from the Earth, star, and exoplanet (black), alongside the same flux degraded to match the desired exposure signal-to-noise ratio (S/N) (red). The second row illustrates the noiseless spectral matrix. Darker regions correspond to stronger absorptions, mainly, from the Earth's atmosphere. The third row corresponds to the noisy spectral matrix $F_{\text{sim}}$, masked (empty pixels) in the regions where over $80\%$ of the flux is absorbed. The horizontal stripes correspond to varying exposure qualities due to the throughput variations. The fourth row showcases the -prepared dataset, where all major contributions have been removed and the exoplanet signal, if present, is buried in the noise.}
\label{fig:S1}
\end{figure}

\begin{figure*}[htb!]
\centering
\includegraphics[angle=0, width=120mm]{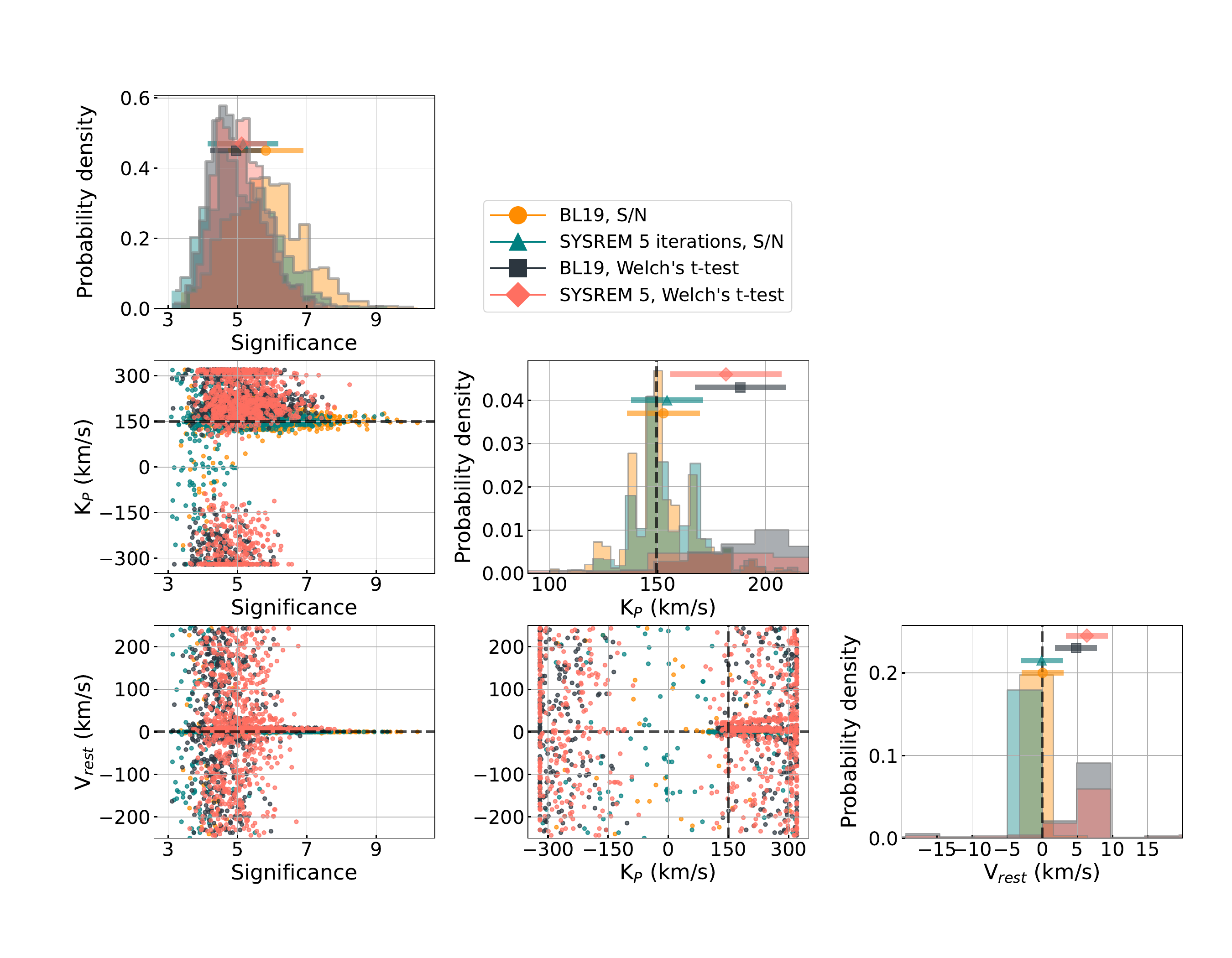}
\caption{Same as Fig.\,\ref{fig:corner_plot} of the main text, but for a cross-correlation step of $3.2$\,km/s, which is the average resolution element of the CARMENES' near-infrared channel.}
\label{fig:S2}
\end{figure*}

\begin{figure}[htb!]
\centering
\includegraphics[angle=0, width=1\columnwidth]{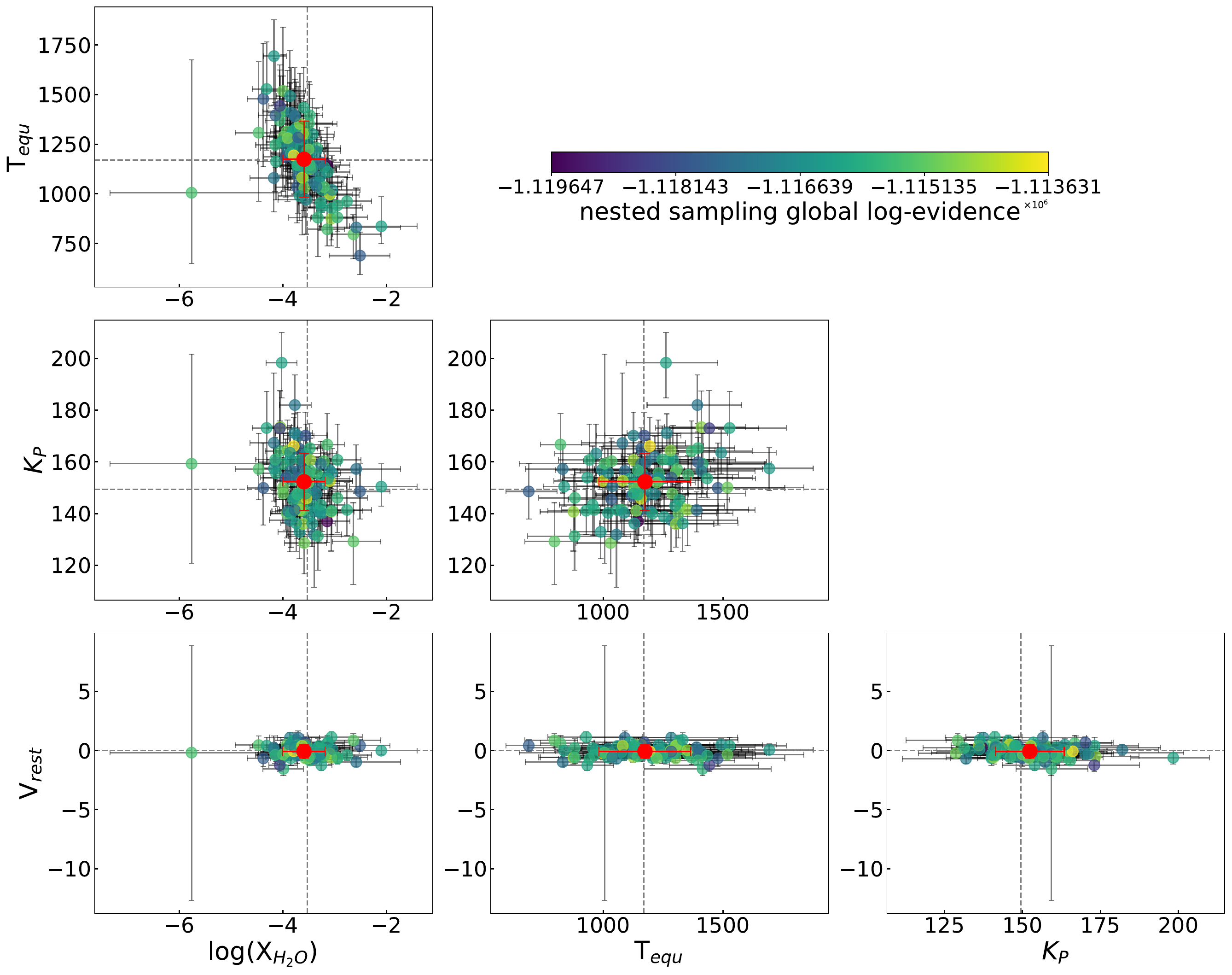}
\caption{
Statistical analysis of $100$ retrievals of key atmospheric parameters for \emph{in-silico} data of HD\,189733\,b. Paired correlations are shown for retrieved water abundances (log(X$_{H_2O}$)), equilibrium temperatures (T$_{equ}$), planetary radial velocity semi-amplitudes ($K_P$), and rest-frame velocities ($V_{rest}$). 1$\sigma$ error bars are indicated. Colors indicate the global evidence of the specific retrieval. Red points and error bars indicate the mean values and uncertainties over all retrievals, and the dashed grey lines represent the true parameter values. These results highlight the consistency of retrievals finding signals under varying conditions, albeit with significant noise-induced variability.}
\label{fig:S3}
\end{figure}

\begin{figure}[htb!]
\centering
\includegraphics[angle=0, width=1.\columnwidth]{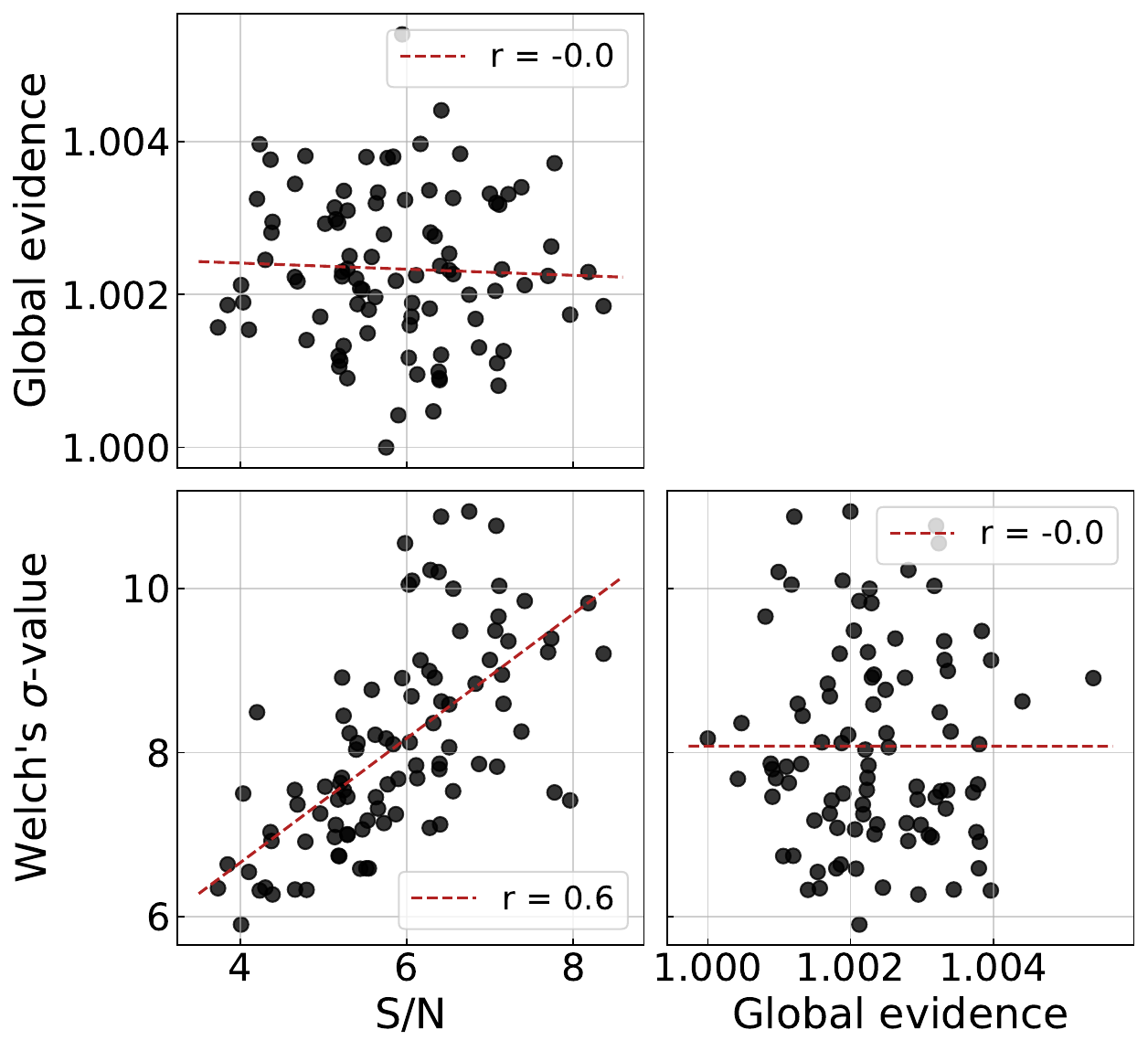}
\caption{Study of detection-significances in $100$ \emph{in-silico} datasets containing an H$_2$O exo-atmospheric signal by using the BL19 preparing-pipeline. The signal-to-noise ratio metric (S/N), Welch's t-test ($\sigma$-values) and the Bayesian-retrieval's global evidence metrics are used to assess the presence and significance of the potential cross-correlation signals in the same datasets. In the scatter-plots, each dot corresponds to a simulated observation. We observe a direct correlation ($r = 0.6$) between the significances obtained by the S/N and Welch's t-test, which is absent when comparing them with the retrievals. This indicates that the datasets offering the highest S/N should also yield the highest Welch's $\sigma$-values, but not necessarily the tightest parameter-constrains. As expected from the results discussed in the main text, the baseline for Welch's $\sigma$-values is notably higher than for the S/N results, resulting in overall higher significance-numbers in the former. Discarded datasets where CCF-techniques fail to disentangle an exoplanet signal, may still reveal useful information by performing a Bayesian retrieval.
}
\label{fig:S4}
\end{figure}

\begin{figure*}[htb!]
\centering
\includegraphics[angle=0, width=120mm]{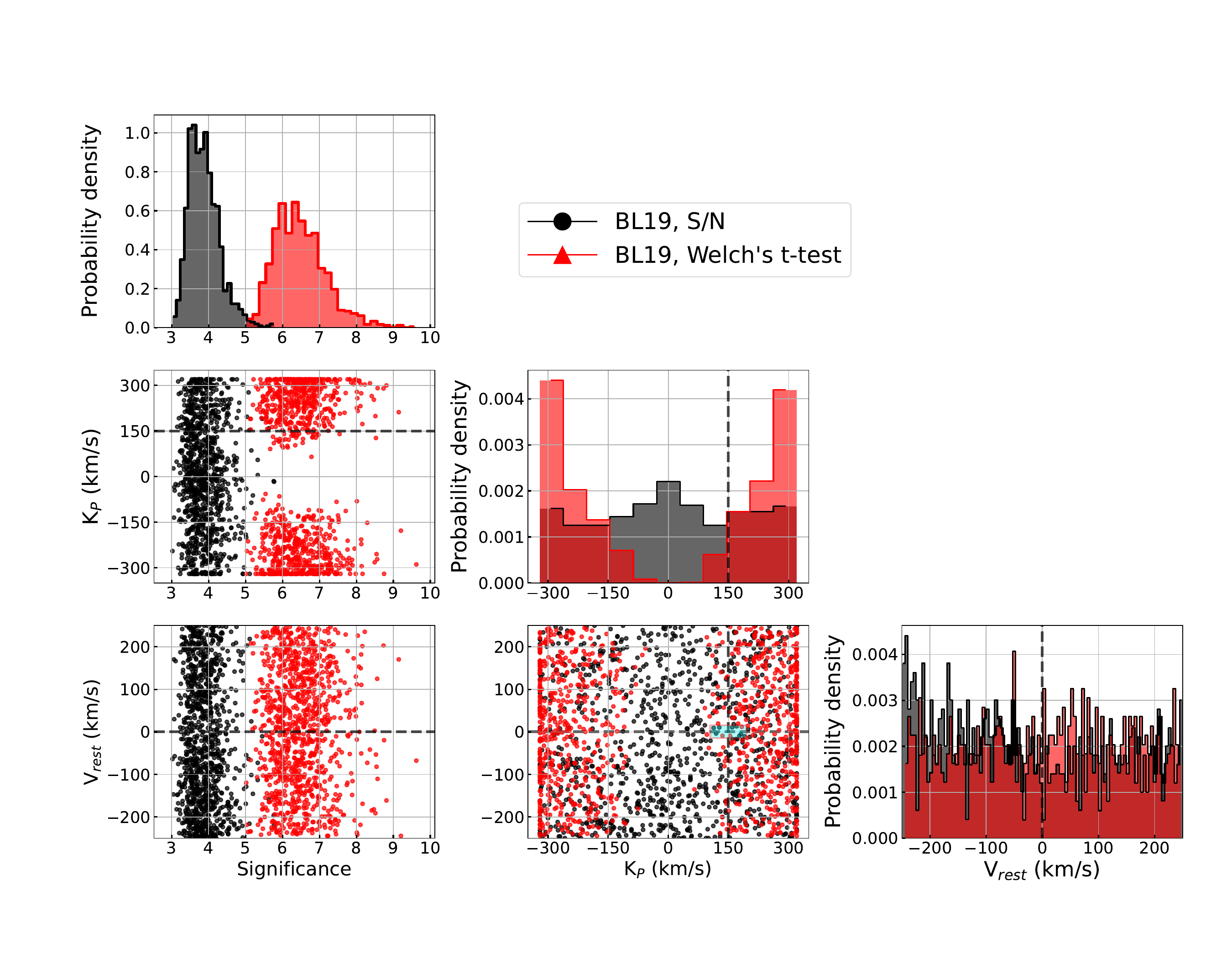}
\caption{Same as Fig.\,\ref{fig:corner_plot} of the main text, but in the case of searching for water vapor in a completely dry exo-atmosphere in $1000$ \emph{in-silico} observations. Results shown for the BL19 preparing pipeline and the signal-to-noise ratio metric. A cross-correlation step of $1.3$\,km/s was used for illustration purposes, due to the appreciable shift between both metrics. The maximum significance signals found are randomly distributed in the parameter space, as expected. The region shaded in cyan around the ground-truth $K_P$\,--\,$V_{rest}$ values shows that only $0.8\%$ of the realizations using the S/N metric (only $0.4\%$ for the Welch's t-test) created noise structures that can be mistaken for real exo-atmospheric H$_2$O signals at these velocities (see discussion in Sect.\,\ref{sect:stats_signal_detections} of the main text).}
\label{fig:S5}
\end{figure*}

\begin{figure*}[htb!]
\centering
\includegraphics[width=1\columnwidth]{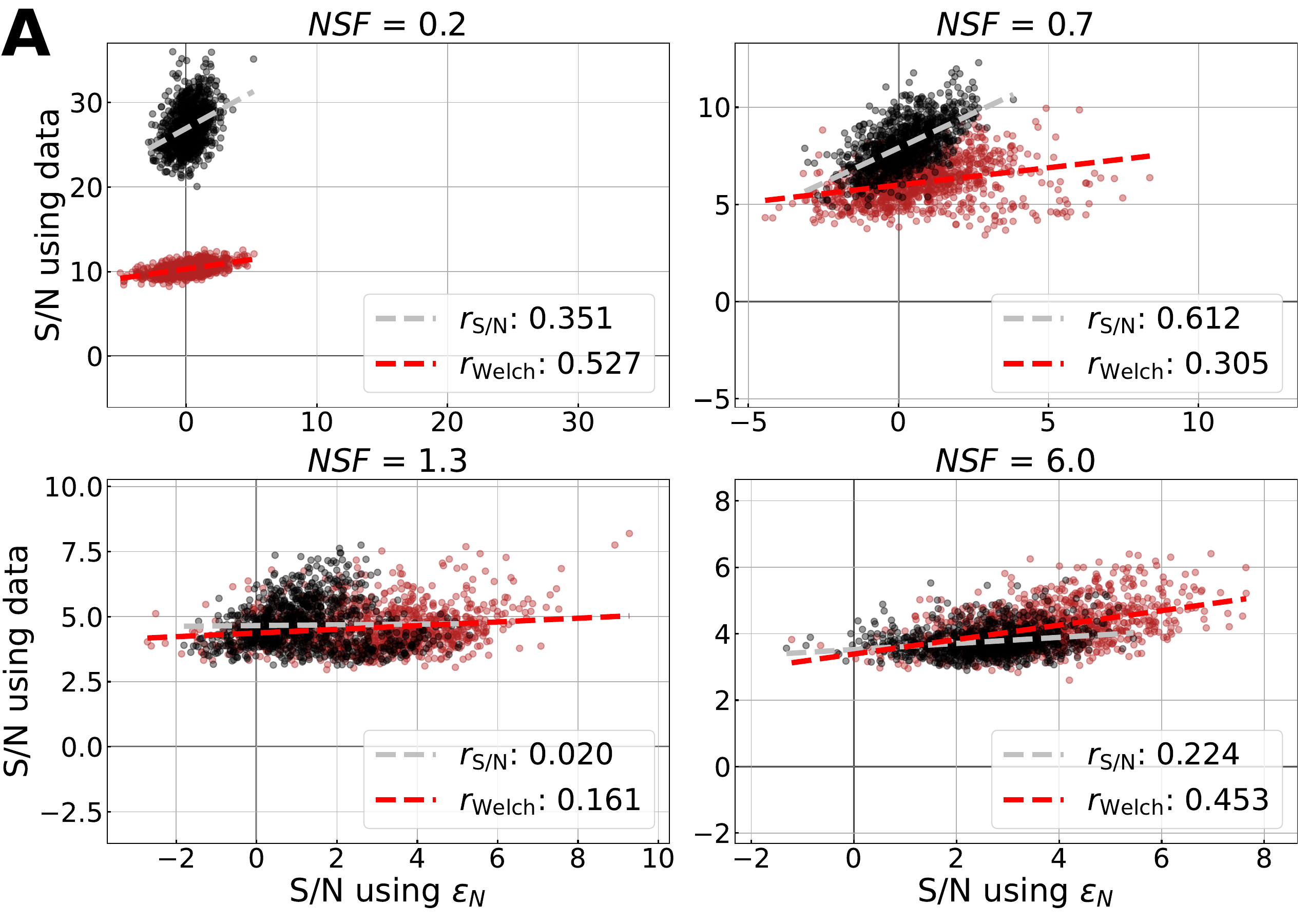}\includegraphics[width=1\columnwidth]{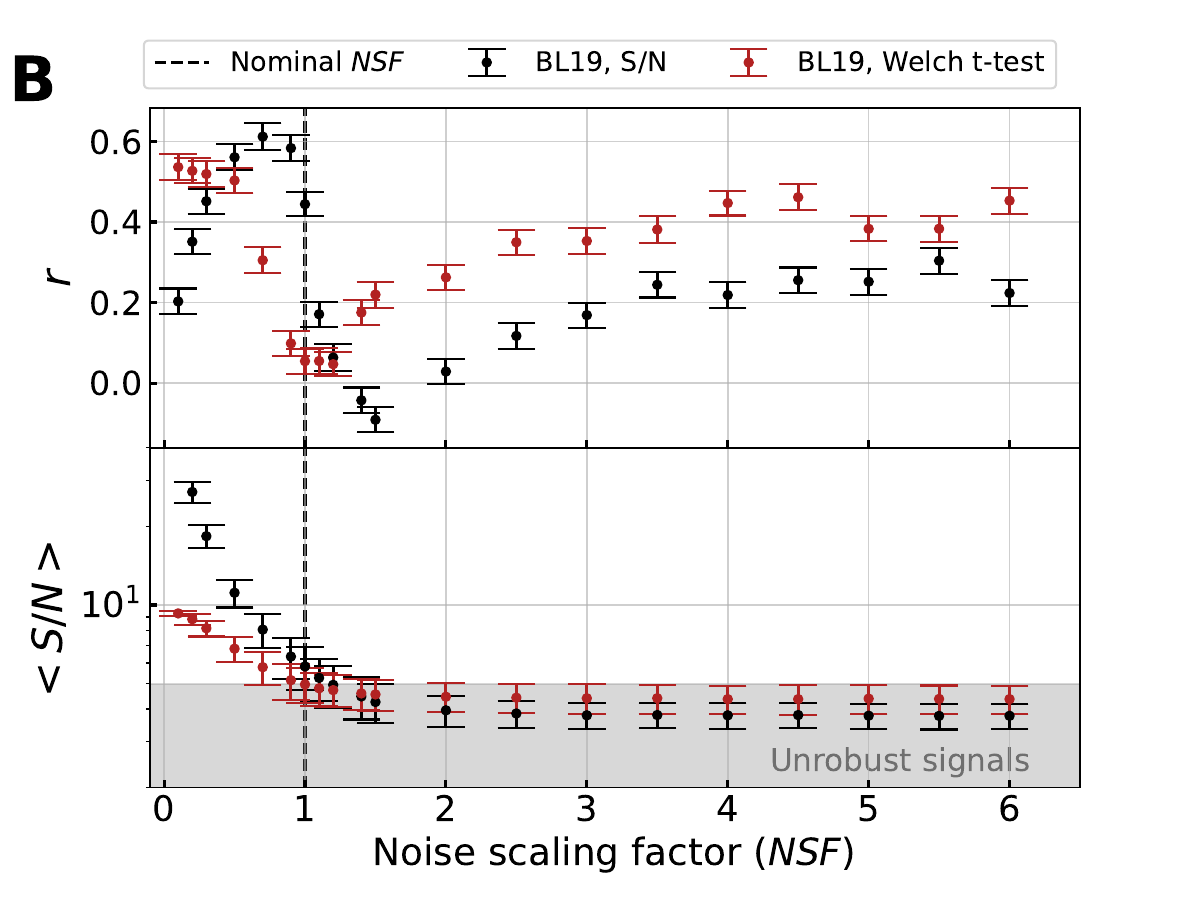}
\hspace{0.01\textwidth}
\caption{Same as Fig.\,\ref{fig:Stat_scale_factors} of the main text, but for a CCF step of $3.2$\,km/s. The behavior in the top panel is similar to the other cases, but the higher basal values of the Welch's t-test metric observed in the bottom panel of Fig.\,\ref{fig:Stat_scale_factors} have disappeared as the in-trail distribution of CCF values is no longer oversampled.}
\label{fig:S6}
\end{figure*}

\begin{figure*}[htb!]
\centering
\includegraphics[width=1\columnwidth]{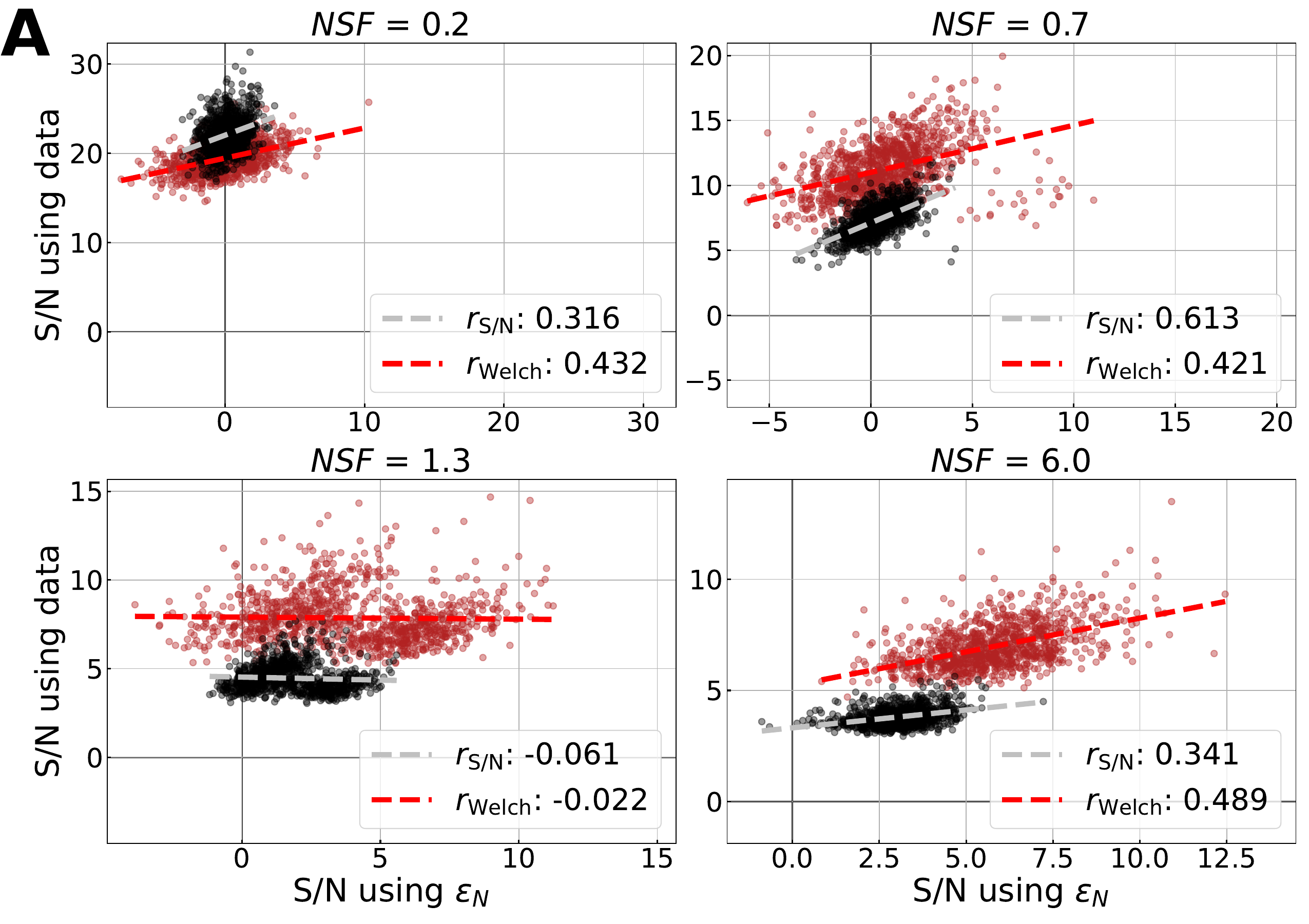}\includegraphics[width=1\columnwidth]{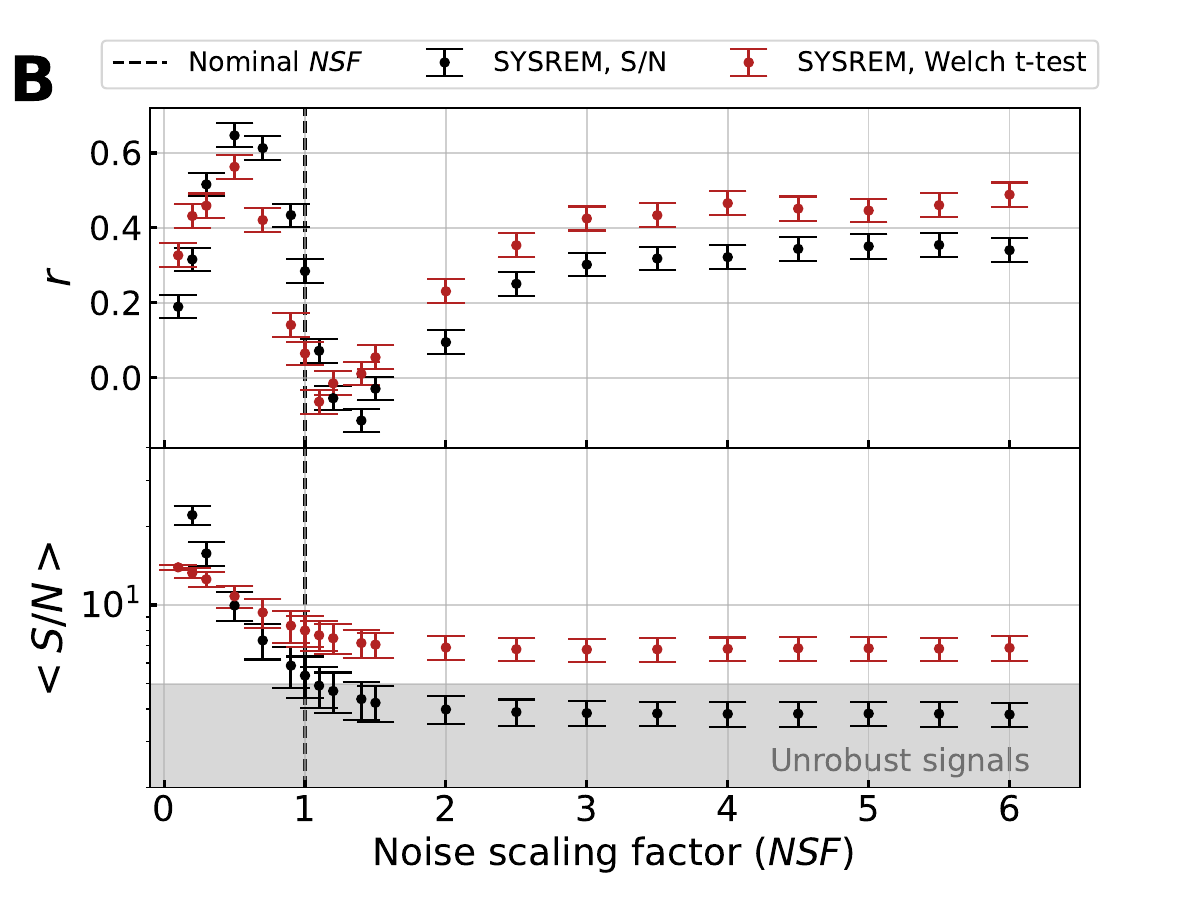}
\hspace{0.01\textwidth}
\caption{Same as Fig.\,\ref{fig:Stat_scale_factors} of the main text, but for the {\tt SYSREM} preparing pipeline, showing a similar behavior as for the BL19 pipeline.}
\label{fig:S7}
\end{figure*}

\begin{figure}[htb!]
\centering
\includegraphics[angle=0, width=1\columnwidth]{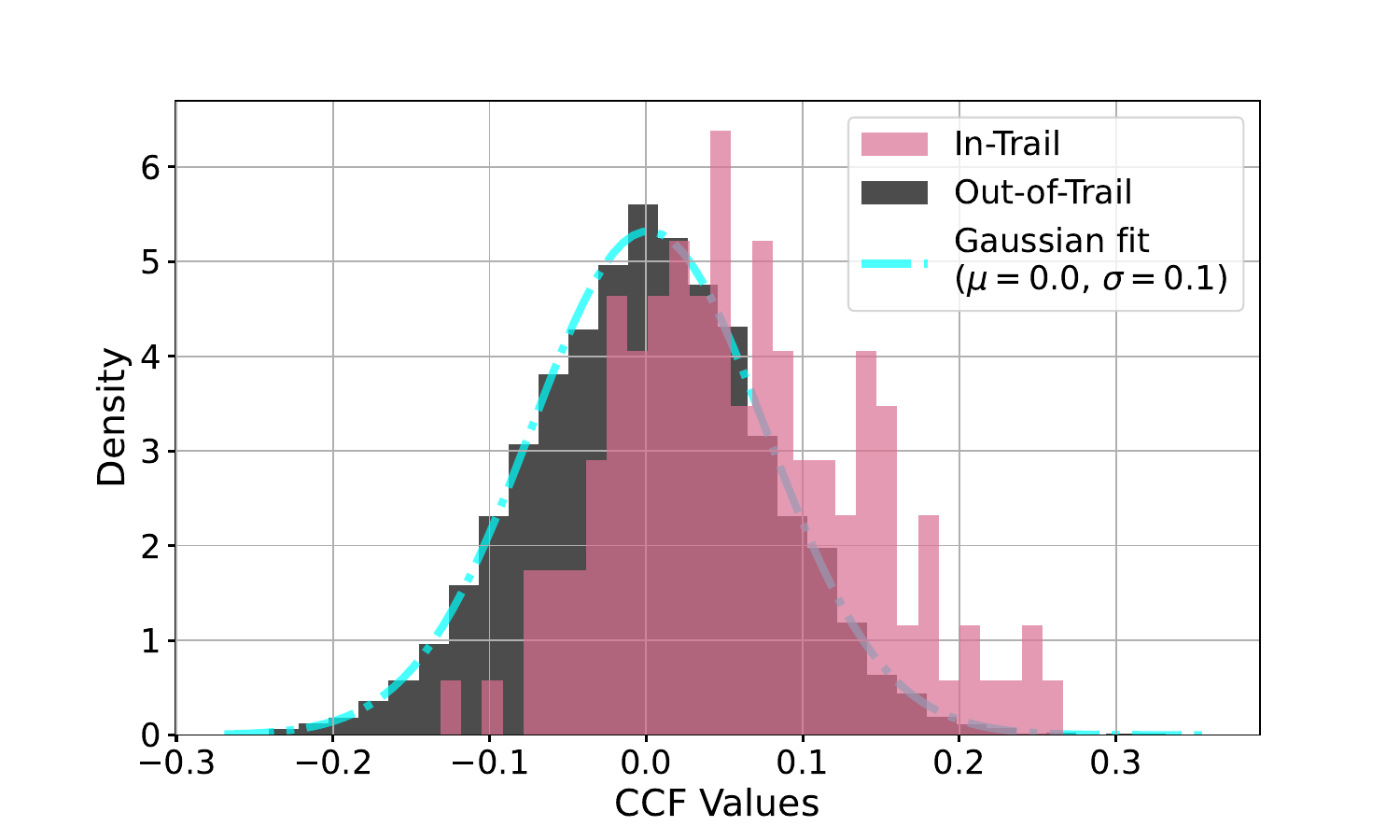}
\caption{Distributions of cross-correlation values (CC values) shown for different regions of the cross-correlation maps. In magenta, CC values around the expected $K_P$ of HD\,189733\,b (in-trail distribution). The rest of the CC values shown in black (out-of-trail distribution) are far from the in-trail signal, since they are obtained by excluding the in-trail points and a $\pm 25$\,km/s region around them. The out-of-trail points are compared with a Gaussian distribution having their same mean and variance (dashed cyan). When a strong exo-atmospheric signal is present in the data, as it is the case here, the in-trail and out-of-trail distributions exhibit notably distinct means.}
\label{fig:S8}
\end{figure}

\begin{figure}[htb!]
\centering
\includegraphics[angle=0, width=1\columnwidth]{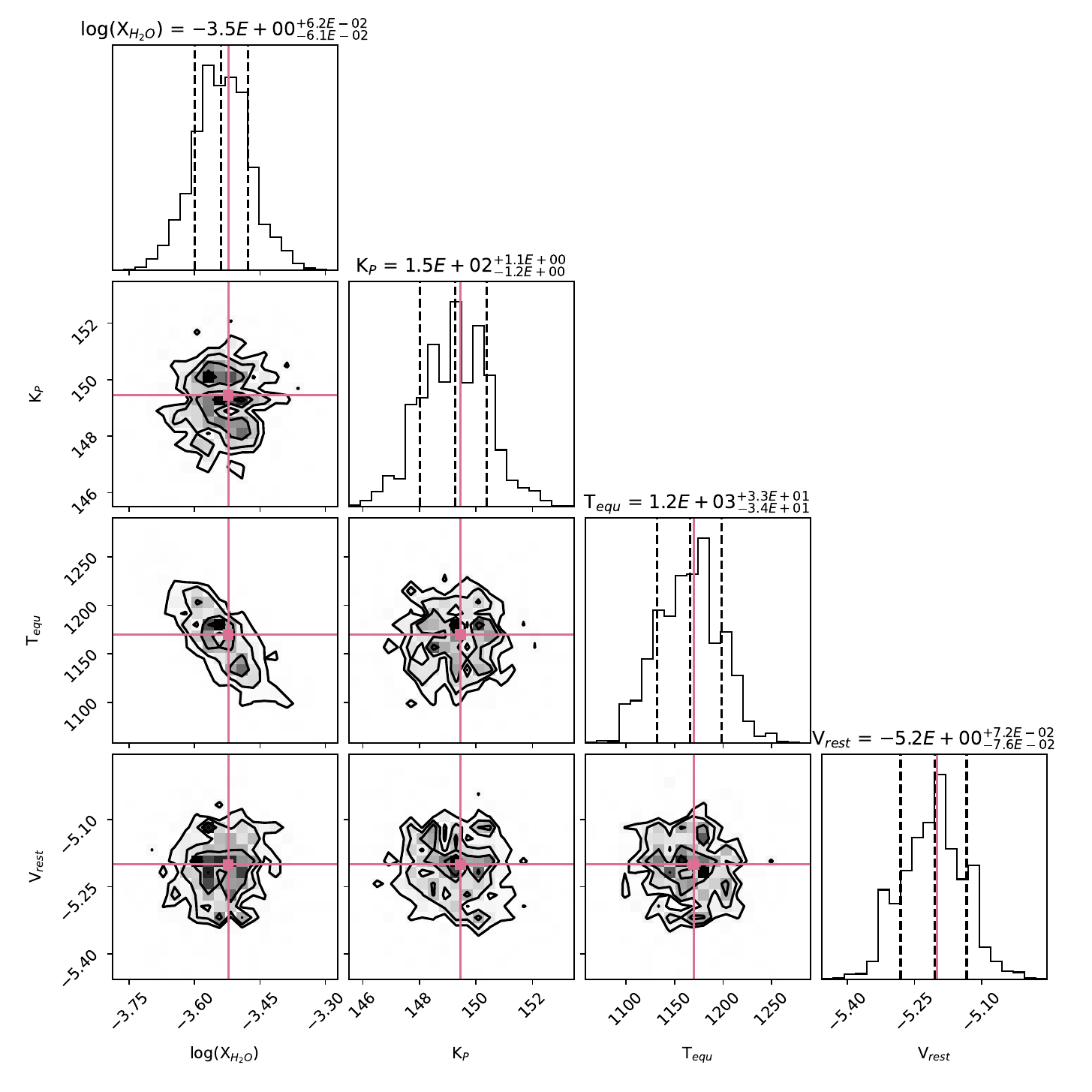}
\caption{Bayesian retrievals in a noiseless dataset of the H$_2$O mass fraction ($\log \left( \text{X}_{\text{H}_2\text{O}} \right)$), of the exoplanet's $K_P$, of the rest-frame velocity ($\text{v}_{\text{rest}}$) of potential signals, and of the equilibrium temperature of the exoplanet (T$_{eq}$). The noiseless retrieval successfully constraints the desired parameters within reasonable uncertainties, which constitutes a solid foundation for our framework.}
\label{fig:S9}
\end{figure}

\end{document}